\documentstyle[12pt, epsf]{article}

\def \M{\Lambda}

\def \f{\tilde{f}}
\def \P{\Psi}
\def \spin{{\rm spin}}
\def \matrix{{\rm matrix}}
\def\s{\sigma}
\def\g{\gamma}
\def \tsalt{\tilde{\mathit F}_{\Lambda}}
\def \tsaltone{\tilde{\mathit F}_1'}
\def \gl{gl_{+\infty}}
\def \m#1{$#1$}

\newcommand{\beq}{\begin{eqnarray}}
\newcommand{\eeq}{\end{eqnarray}}
\newtheorem{lemma}{Lemma}
\begin{document}
\begin{center}
   {\LARGE\bf A Lie Algebra for Closed Strings, Spin Chains and Gauge Theories } \\
   \vspace{1cm}
   {\large\bf C.-W. H. Lee and  S. G. Rajeev} \\
   {\em Department of Physics and Astronomy, University of Rochester, 
    Rochester, New York 14627} \\
   \vspace{.3cm}
   {May 30th, 1998.} \\
   \vspace{1cm}
   {\large\bf Abstract}
\end{center}

We consider quantum dynamical systems whose degrees of freedom are
described by \m{N\times N} matrices, in the planar limit  \m{N\to
\infty}. Examples are gauge theories and the M(atrix)-theory of strings.
States  invariant  under \m{U(N)} are 
`closed strings',  modelled by traces of products of 
matrices. We have discovered that the 
$U(N)$-invariant operators acting on both open and closed string states form a
remarkable new Lie algebra which we will call the heterix algebra.
(The simplest special case, with one degree
of freedom, is an extension of the Virasoro algebra by the
infinite-dimensional general linear algebra.)  Furthermore, these operators
acting on closed string states only form a quotient algebra of the heterix
algebra.  We will call this quotient algebra the cyclix algebra.  
We express the Hamiltonian of some gauge field
theories (like those with adjoint matter fields and dimensionally
reduced pure QCD models) as elements of this Lie algebra.
Finally, we apply this cyclix
algebra to establish an isomorphism between certain planar matrix
models and quantum spin chain systems. Thus  we obtain some matrix
models solvable in the planar limit; e.g., matrix models associated
with the  Ising model, the XYZ model, models satisfying the
Dolan-Grady  condition and the chiral
Potts model.  Thus our cyclix  Lie algebra describes  the dynamical symmetries of
quantum spin chain systems, large-$N$ gauge field theories, and the 
M(atrix)-theory of strings.  

\begin{flushleft}
\end{flushleft}
\pagebreak

\section{Introduction}

The string is emerging as a verstile concept in physics, second only
to the notion of  a particle in its usefulness. The instantaneous
configuration of a   string can be 
thought of roughly as a  curve in space, with an energy 
proportional to its length. The curve may be closed,  or open with some
extra degrees of freedom stuck at its endpoints.  The string is being
intensely studied as the fundamental object in the quantum theory
of gravity and perhaps even the unified theory of all forces of nature.  

The modern notion of a relativistic string theory originated in attempts
to understand hadron dynamics, in the sixties and early
seventies. However, it is now established that Quantum Chromodynamics (QCD) 
is the fundamental theory of strong interactions: the hadrons are
not elementary particles but instead are bound states of quarks and
gluons.  In spite of this, attempts to understand hadron
dynamics in terms of strings have not altogether ceased.  Many
features of hadron dynamics which originally prompted people to construct the
string theory still do not have a satisfactory explanation. 
For example the squares of the masses of hadrons increase linearly
with their angular momenta, a  characteristic property of strings.
Indeed, it seems very plausible that there is a version of string theory which
is equivalent to QCD but that directly describes hadrons.  (Such a
theory has been constructed in two dimensions \cite{rajeev}.)  Since the gluon
degrees of freedom are represented by matrices, this suggests that at
least this type of  string can be thought of as made of more
elementary entities described  by products of matrices.  Connections
to  non-commutative geometry are also natural as suggested in
Ref.\cite{uym}.  

Meanwhile in recent years the  strings  that appear in  theories of quantum
gravity have also come to be seen as made of more elemental objects
(`string bits' \cite{th79, thorn96}) again described by matrices
\cite{susskind, kawai}. Indeed the underlying mathematical structures
have much in common with earlier work on QCD with some extra
symmetries (such as supersymmetry) added. These M(atrix)-theories are
thought to  describe not only string-like excitations but also
M(embranes)  and are collectively known as the M-theory.  The
underlying  symmetries and
geometrical significance of this theory still remain M(ysterious).

Both of the above applications of string theory involve fundamental
degrees of freedom represented by  \m{N\times N} matrices. In both
cases the limit \m{N\to \infty} is important, as it  simplifies many
issues. There are in fact many inequivalent ways of letting \m{N} go
to  infinity, the simplest one is the so-called planar large-\m{N}
limit. It is called as such because it was 
originally \cite{thooft}  constructed by summing Feynman
diagrams which have planar topology. (Other limits such as the double
scaling limit which gives the continuum string theory are of interest as
well.)  We view this planar limit as a linear approximation to the 
general large-$N$ limit discussed in Ref.~\cite{rajeevturgut}.

Thus, the  understanding of the quantum matrix models  is of
fundamental importance to  modern theoretical physics. The two great
 challenges of theoretical physics ---
that of understanding hadron dynamics and that of developing a quantum
theory of gravity --- are both intertwined with the dynamics of quantum
matrix models.

We have  discovered the fundamental dynamical symmetry of matrix models in the
large-\m{N} limit: it is  a remarkable new Lie algebra which we will
call the heterix algebra. It is thus the algebra of symmetries that
will help us solve both of the two great problems of theoretical
physics. In addition,  we will find   a connection to a third
branch of theoretical physics: we can understand
the solvability of quantum spin chain models within the formalism of
the heterix algebra. 

The outline of this paper is as follows.  In Section~\ref{s5}, we will introduce heterix algebra axiomatically.  We 
will also discuss aspects of its structure such as its  Cartan subalgebra and root vectors.
In particular, we will show that this Lie algebra is an extension of another Lie algebra by 
$gl_{+\infty}$ \cite{kac}.  In Section~\ref{s6}, we will study the simplest heterix algebra in which there is only 
one degree of freedom.  It will turn out that the non-commutative algebra is simplified to the Virasoro algebra.  
Thus our algebra can be viewed as a generalization of the Virasoro algebra.   In Section~\ref{s2}, we will show 
that gauge invariant observables in the planar large-$N$ limit can be formulated in terms of the elements of this 
algebra.  Moreover, closed string bit states in the this large-$N$ limit provide a representation for this Lie 
algebra.  However, this representation is not faithful and thus there are many relations among some elements of 
this algebra.  We will quotient out these relations and get another Lie algebra --- the cyclix algebra 
\cite{leerajlett}.

In Section~\ref{s9}, we will discuss the applications of the cyclix algebra to various gauge field theories, which,
in the planar large-$N$ limit, are multi-matrix models.  In Section~\ref{s7}, we will construct more tractable
multi-matrix models with the aid of quantum spin chain models.  In addition,  we will find that we can understand
the solvability of quantum spin chain models within the formalism of the cyclix algebra.

\section{The Heterix Lie Algebra}
\label{s5}

In this section we will give a self-contained definition of a certain
Lie algebra, which we will call the heterix algebra.  
In a latter section we will see that closed string
operators provide a representation for this algebra, albeit an
unfaithful one. The defining representation in this section is on the
combined state space of open and closed strings and is, of course,
faithful. 

\begin{figure}
\epsfxsize=5in
\centerline{\epsfbox{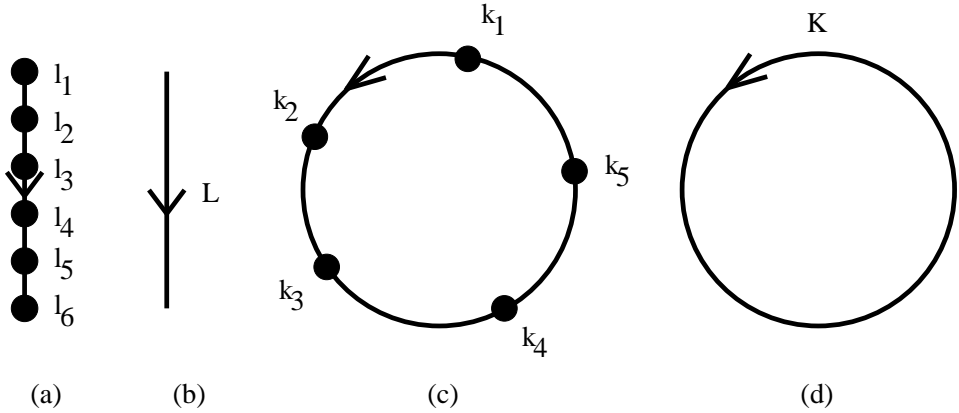}}
\caption{\em (a) A typical open string state $s^L$ in detail.  There is a series of 6 indices $l_1$, $l_2$,
\ldots, and $l_6$.  Each index is represented by a solid bead.  
These solid beads are connected by thick lines (in contrast with the thin lines in latter figures).
The arrow indicates the direction of the integer sequence $L$.
(b) An open string state in brief.  The beads are omitted.  They will be consistently omitted in all ensuing brief
figures.  The integer sequence is represented by the letter $L$.  The length of the line does not indicate the
number of indices the line carries.
(c) A typical closed string state $\Psi^{(K)}$ with 5 indices in detail.  There is a 
series of indices $k_1$, $k_2$, \ldots, and $k_5$.  Notice the cyclic symmetry of this state.  
(d) A closed string state in brief.  The state is represented by the equivalence class $(K)$, but in the
figure we simply label the state as $K$.  The size of the
circle does not indicate the number of indices the circle carries.}
\label{f1}
\end{figure}

Define a complex vector space \m{{\cal T}_{op}} of {\em finite} linear
combinations of orthogonal states\footnote{See Appendix~\ref{s1-1} for a more
detailed explanation of the notation.} of an open string:
\m{s^L=s^{l_1l_2\cdots l_d}}.  This is just the space of
all tensors (with no particular symmetry properties) on the complex
vector space \m{C^{\Lambda}}:
\beq
	{\cal T}_{op}=\oplus_{r=1}^\infty \otimes^r C^{\Lambda}.
\eeq
This vector space should be thought of as  the space of states of an open string
propagating in a discrete model of space-time with \m{\Lambda} points.
In addition define \m{{\cal T}_c} of finite linear combinations of
orthogonal states \m{\P^{(K)}=\P^{(k_1\cdots k_c)}} labelled by cyclically symmetric indices.
 Thus \m{{\cal T}_c} is the space of {\it cyclically symmetric
tensors} on \m{C^\Lambda} and should be viewed
as the space of states of a closed string propagating in  discretized
space. Our Lie algebra
will be defined by the commutators of some operators on the total space of
closed and open strings \m{{\cal T}={\cal T}_c\oplus{\cal T}_{op}}.  
Thus \m{\cal T} has an orthogonal basis
\m{\{\P^{(K)}, s^L \}} as \m{(K)} ranges over (non-empty) equivalence  classes of cyclically
symmetric sequences and \m{L} over all non-empty sequences. 
Typical open and closed string states are illustrated in Fig.~\ref{f1}. (The reader can compare Figs.~\ref{f1}(a) 
and (b) with Figs.~\ref{f1}(a) and (b) in an accompanying paper \cite{opstal}.)

\begin{figure}
\epsfxsize=3.5in
\centerline{\epsfbox{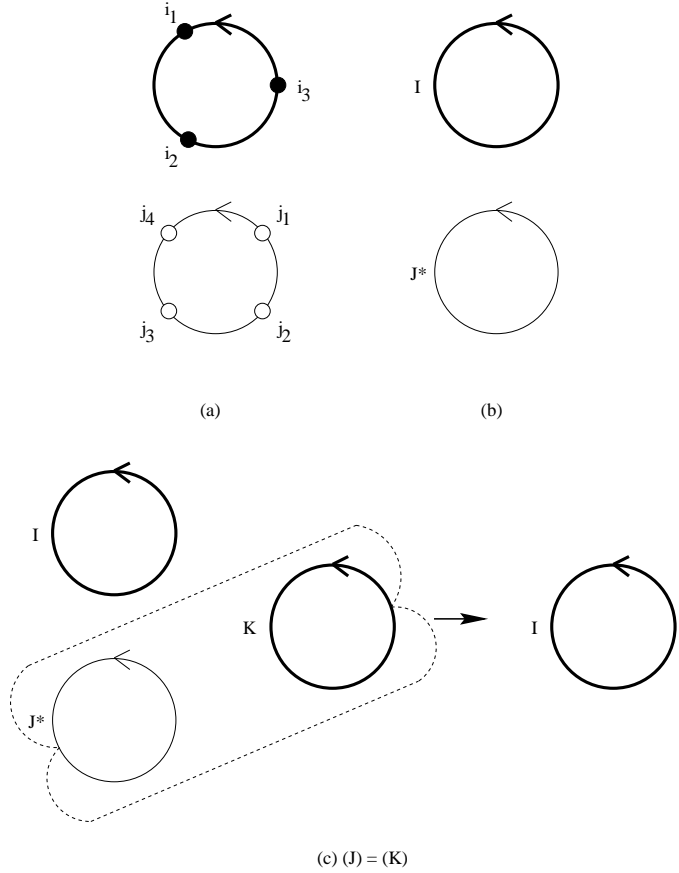}}
\caption{\em (a) A detailed diagrammatic representation of a typical $\f^{(I)}_{(J)}$.  The upper circle represents 
$(I)$,  whereas the lower circle represents $(J)$.  In this example, there are 3 indices in $I$ and 4 indices in
$J$.  An index in $I$ is represented by a solid bead, whereas an 
index in $J$ is represented by a hollow bead.  Moreover, the solid beads are joined by thick lines, while
the hollow ones are joined by thin lines.  Note the sequence $J$ is put in the clockwise instead of the 
anti-clockwise direction.  (b) $\f^{(I)}_{(J)}$ in brief.  An asterisk is added to $J$ to signify the fact
that $J$ is in the anti-clockwise direction.  (c) The 
action of $\f^{(I)}_{(J)}$ on $\P^{(K)}$.  The dotted lines connect the line segments to be `annihilated' together.
The figure on the right of the arrow is the resultant closed string state.}
\label{f5.2}
\end{figure}

Now let us introduce some  operators that act on these states.  Define the operator $\f^{(I)}_{(J)}$ by its action 
on ${\cal T}$ as follows: 
\begin{equation}
   \f^{(I)}_{(J)} \P^{(K)} = \delta^K_{(J)} \P^{(I)}\; \mbox{and}\; 
   \f^{(I)}_{(J)} s^L = 0. 
\label{5.3}      
\end{equation}
This operator just converts the closed string state 
\m{\P^{(K)}} to \m{\P^{(I)}}, with a multiplicative factor which is the
number of cyclic permutations of \m{K} that is equal to \m{J}. It gives
zero if there are no such permutations or if it acts on an open string state. 
Clearly \m{\f^{(I)}_{(J)}} is a finite rank operator.
The operator $\f^{(I)}_{(J)}$ is shown pictorially in Fig.~\ref{f5.2}.
Also shown in the same 
figure is the action of $\f^{(I)}_{(J)}$ on $\P^{(K)}$.

In addition let us define the operator \m{\gamma^I_J} by its actions on closed strings --- 
\begin{equation}
   \gamma^I_J \P^{(K)} = \delta^K_{(J)} \P^{(I)} + \sum_{K_1 K_2 = (K)} \delta^{K_1}_J \P^{(I K_2)} 
\label{5.1}      
\end{equation}
--- and on open strings as shown:
\begin{equation}
   \gamma^I_J s^L = \delta^L_J s^I + \sum_{L_1 L_2 = L} \delta^{L_1}_J s^{I L_2} 
   + \sum_{L_1 L_2 = L} \delta^{L_2}_J s^{L_1 I} + \sum_{L_1 L_2 L_3 = L} \delta^{L_2}_J s^{L_1 I L_3}.
\label{8.1}    
\end{equation}
Clearly, it is the direct sum of an operator \m{g^I_J} acting on closed strings and \m{\s^I_J} acting on open
strings: \m{\gamma^I_J = g^I_J \oplus \s^I_J}. $\s^I_J$ has been defined in a previous paper \cite{opstal}.  These 
operators are not finite rank, unlike the \m{\f^{(I)}_{(J)}}; indeed they are not even bounded.

Let us describe Eq.(\ref{5.1}) in words.  Regard $(K)$ as a circle of
points labelled by positive integers. If $J$ has more indices than $K$, then
the action of \m{\gamma^I_J} on \m{\P^{(K)}} gives zero. 
When $J$ is shorter than $K$, 
each  segment of \m{K} that is identical to $J$, is  replaced with $I$
and then a sum over all the resultant terms is taken.
This is done even if some of the  segments agreeing with \m{J} 
overlap partially with each other.  If \m{J} does not overlap with any
segment of \m{K}, then we get zero.

Next, consider the case when the length of $J$ 
is the same as that of $K$,
i.e., $\#(J) = \#(K) = c$. Then we get  zero if no cyclic permutaion
of \m{J} agrees with \m{K}; otherwise, we replace \m{K} with \m{I} and
multiply $\Phi^{(I)}$ by the number of cyclic permutataions of \m{K}
that agree with \m{J}.  

Eq.({\ref{5.1}) is illustrated in Fig.~\ref{f5.1}.  Using the identities in 
Appendix~\ref{s1-1}, Eq.(\ref{5.1}) can be rewritten as
\begin{equation}
   \g^I_J \P^{(K)} = \delta^K_{(J)} \P^{(I)} + \sum_A \delta^K_{(J A)} \P^{(I A)}.
\label{5.1.1}
\end{equation}   
This form is more convenient for some calculations.

We can understand Eq.(\ref{8.1}) analogously. Again, if \m{J} is longer  than
\m{K}, we get zero. Otherwise, each segment of
\m{K} that agrees with \m{J} is replaced with \m{I} and a sum over all
such resultants  is taken. The first term represents the situation
when \m{J} is identical to \m{K}; the second when J agrees with the
beginning of \m{K}; the third when \m{J} agrees with the end of \m{K};
the last term, which is the `generic' one, describes the situation
when \m{J} agrees with some segment in the middle of \m{K}. Of course
we get zero if there is no segment of \m{K} that agrees with \m{J}.
Eq.(\ref{8.1}) is depicted in Fig.~\ref{5.1.1}.

The reader can notice from the figures an analogy between the action
of our operators on states  and the action of some  viruses  on 
 DNA molecules.

\begin{figure}
\epsfxsize=4.5in
\centerline{\epsfbox{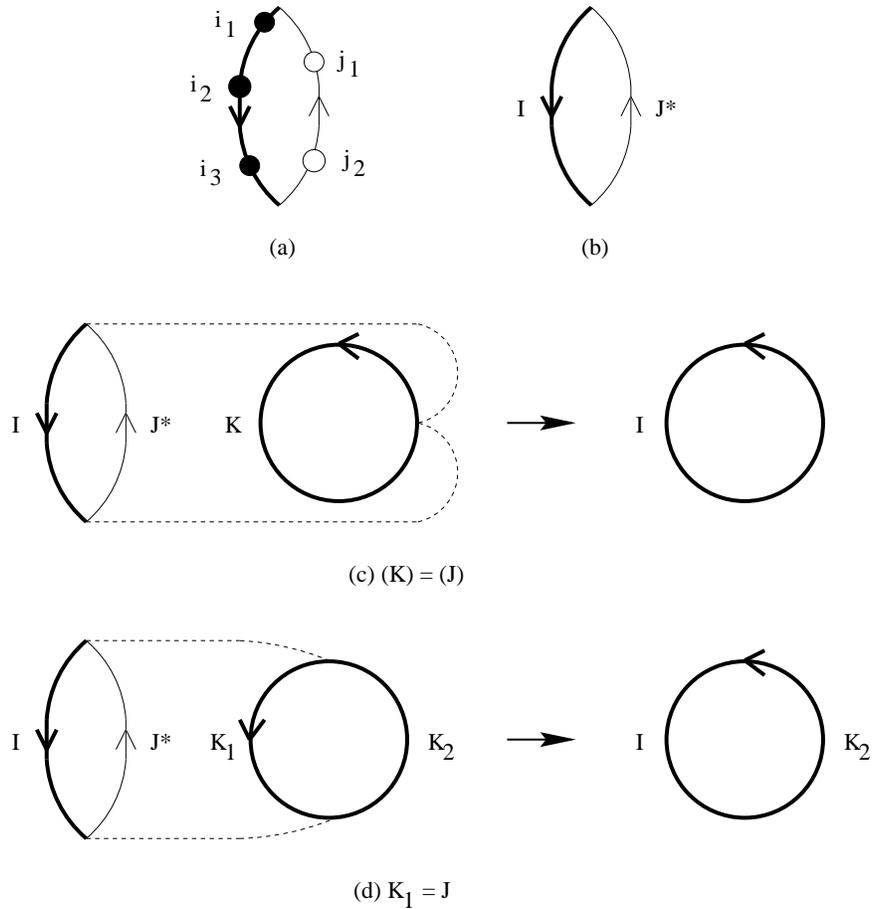}}
\caption{\em (a) A typical $\gamma^I_J$.  There are 3 indices in $I$ and 2 indices in $J$.  Note that $J$ is in
reverse.  (b) A concise representation of $\gamma^I_J$.  Again an asterisk is added to $J$ to signify the fact
that $J$ is put in reverse.  (c) and (d)  The action of $\gamma^I_J$ on $\P^{(K)}$.  Only the first two terms of
Eq.(\ref{5.1}) are shown.} 
\label{f5.1}
\end{figure}

\begin{figure}[ht]
\epsfxsize=4.5in
\centerline{\epsfbox{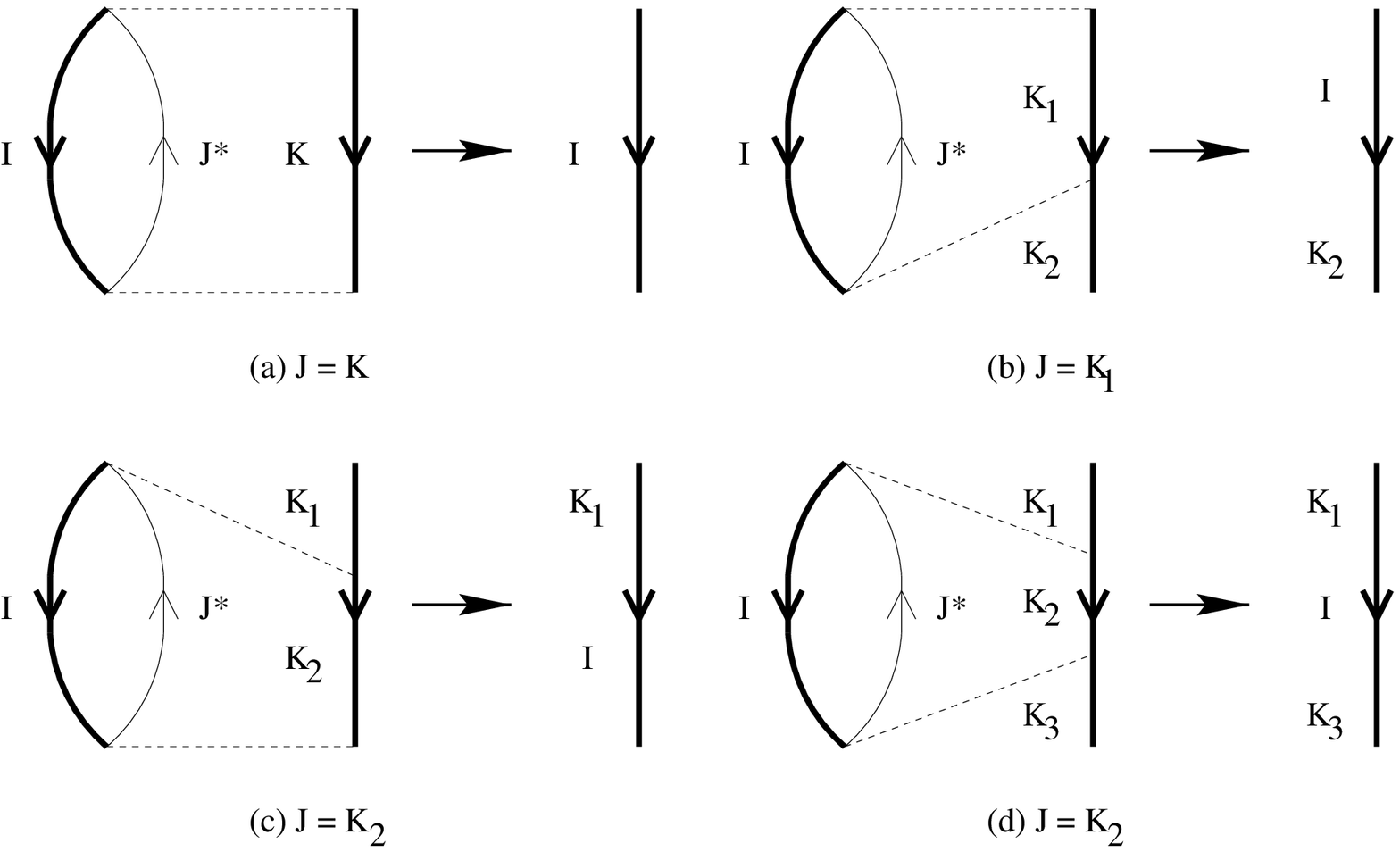}}
\caption{\em (a) to (d)  The action of $\gamma^I_J$ on $s^L$ (c.f. Fig~8 in an accompanying paper \cite{opstal}).  
Only the first four terms are shown.} 
\label{f5.1.1}
\end{figure}

Altogether, \m{\{\f^{(I)}_{(J)},\g^I_J\}} is
a linearly independent set of operators.  This fact, the proof of which can be found in Appendix~\ref{a1}, is of 
some importance, as we will define a Lie algebra with this set as a basis soon.
The operators \m{g^I_J} acting on closed string states alone do not
form a  linearly independent set,
however.  We will derive some relations among them in a latter section.

The product of two operators \m{\f^{(I)}_{(J)}\f^{(K)}_{(L)}} is a finite linear combination of such operators.  
Moreover, the \m{\f^{(I)}_{(J)}}'s are linearly independent so that they span an
associative algebra \m{\tsalt}. Their commutators span a Lie algebra
which we will also call \m{\tsalt}. 
However the product \m{\gamma^I_J\gamma^K_L} cannot be written in general as a
finite linear combination of the \m{\gamma}'s and $\f$'s.  A proof of this statement can be found in 
Appendix~\ref{a2}. 

Nevertheless, it is a remarkable fact
that the commutators of any two string-like operators defined above can in fact be written as finite linear
combinations of themselves. In fact we see, by a straightforward
(but tedious) computation, that the commutator between
two $\g$'s reads
\begin{eqnarray}
   \lefteqn{ \left[ \g^I_J, \g^K_L \right] = 
   \delta^K_J \g^I_L + \sum_{J_1 J_2 = J} \delta^K_{J_2} 
   \g^I_{J_1 L} + \sum_{K_1 K_2 = K} \delta^{K_1}_J
   \g^{I K_2}_L } \nonumber \\
   & & + \sum_{\begin{array}{l}
		  J_1 J_2 = J \\
		  K_1 K_2 = K
	       \end{array}}
   \delta^{K_1}_{J_2} \g^{I K_2}_{J_1 L} 
   + \sum_{J_1 J_2 = J} \delta^K_{J_1} \g^I_{L J_2}
   + \sum_{K_1 K_2 = J} \delta^{K_2}_J \g^{K_1 I}_L \nonumber \\
   & & + \sum_{\begin{array}{l}
		  J_1 J_2 = J \\
		  K_1 K_2 = K
	       \end{array}}
   \delta^{K_2}_{J_1} \g^{K_1 I}_{L J_2}
   + \sum_{J_1 J_2 J_3 = J} \delta^K_{J_2} \g^I_{J_1 L J_3} 
   + \sum_{K_1 K_2 K_3 = K} \delta^{K_2}_J \g^{K_1 I K_3}_L \nonumber \\
   & & + \sum_{\begin{array}{l}
   		  J_1 J_2 = J \\
   		  K_1 K_2 = K
   	       \end{array}}
   \delta^{K_1}_{J_2} \delta^{K_2}_{J_1} \f^{(I)}_{(L)}
   + \sum_{\begin{array}{l}
   	      J_1 J_2 J_3 = J \\
   	      K_1 K_2 = K
   	   \end{array}}
   \delta^{K_1}_{J_3} \delta^{K_2}_{J_1} \f^{(I)}_{(J_2 L)}
   \nonumber \\
   & & + \sum_{\begin{array}{l}
   		  J_1 J_2 = J \\
   		  K_1 K_2 K_3 = K
   	       \end{array}}
   \delta^{K_1}_{J_2} \delta^{K_3}_{J_1} \f^{(I K_2)}_{(L)} \nonumber \\
   & & + \sum_{\begin{array}{l}
		  J_1 J_2 J_3 = J \\
		  K_1 K_2 K_3 = K
	       \end{array}}
   \delta^{K_1}_{J_3} \delta^{K_3}_{J_1} \f^{(I K_2)}_{(J_2 L)} 
    - (I \leftrightarrow K, J \leftrightarrow L). 
\label{5.5}
\end{eqnarray}
Moreover, the commutators between a $\g$ and an $\f$, and between two $\f$'s, are
\begin{equation}
   \left[ \g^I_J, \f^{(K)}_{(L)} \right] =
   \delta^K_{(J)} \f^{(I)}_{(L)} + \sum_{K_1 K_2 = (K)} \delta^{K_1}_J      
   \f^{(I K_2)}_{(L)} - \delta^I_{(L)} \f^{(K)}_{(J)} -
   \sum_{L_1 L_2 = (L)} \delta^I_{L_2} \f^{(K)}_{(L_1 J)}  
\label{5.6}
\end{equation}
and
\begin{equation} 
   \left[ \f^{(I)}_{(J)}, \f^{(K)}_{(L)} \right] = 
   \delta^K_{(J)} \f^{(I)}_{(L)} -
   \delta^I_{(L)} \f^{(K)}_{(J)}
\label{5.7}
\end{equation}
respectively.  The proof of the above three equations will be relegated to Appendix~\ref{a4}. 
Some terms in Eq.(\ref{5.5}) are depicted in Figs.~\ref{f5.3} to \ref{f5.4.1}.  
Illustrated in Fig.~\ref{f5.5} are the two Lie algebraic relations Eqs.(\ref{5.6}) and (\ref{5.7}).
Let us call the Lie algebra with the basis \m{\{ \f^{(I)}_{(J)}, \g^I_J \} } and the above commutation
relations the {\em heterix algebra}, $\hat{\Gamma}_{\Lambda}$.

\begin{figure}
\epsfxsize=5in
\centerline{\epsfbox{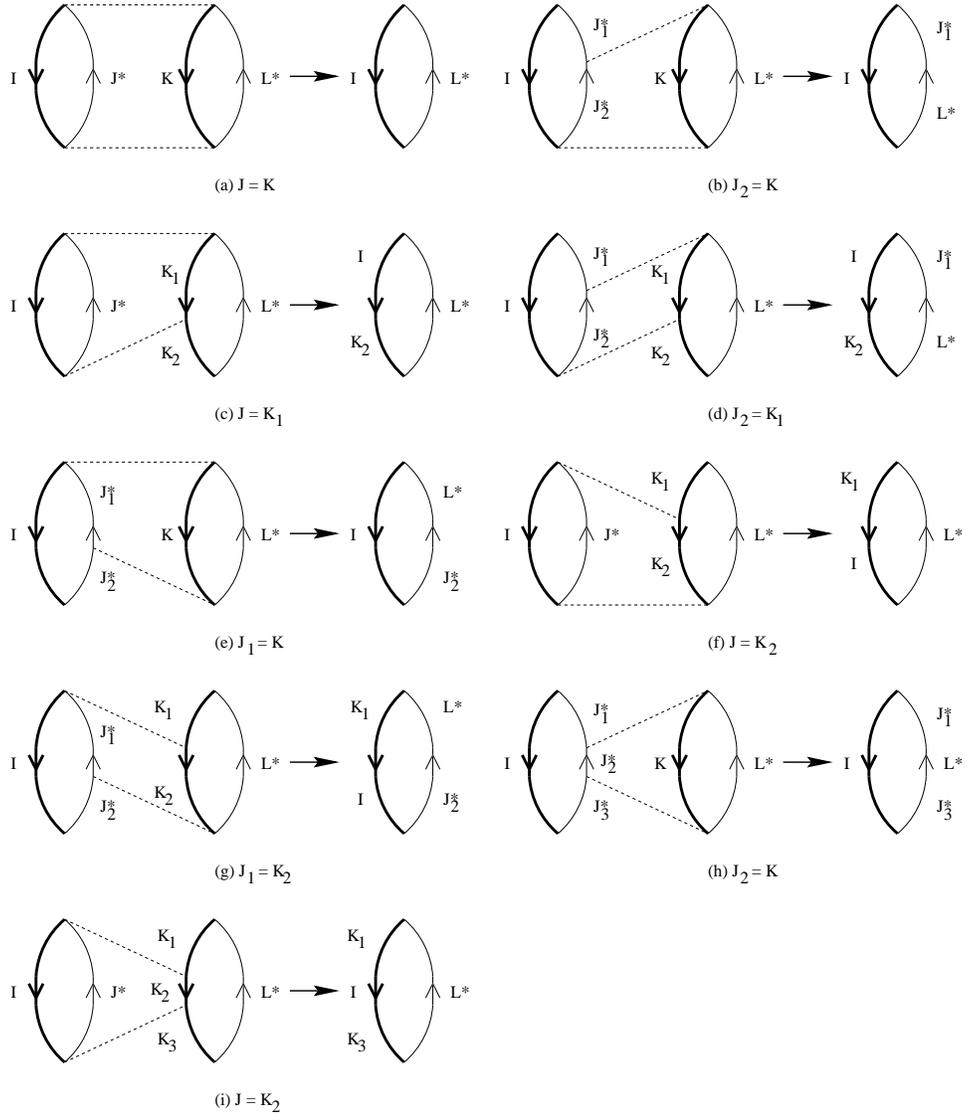}}
\caption{\em Diagrammatic representations of the first 9 terms on the R.H.S. of Eq.(\ref{5.5}).}
\label{f5.3}
\end{figure}

\begin{figure}
\epsfxsize=3.3in
\epsfysize=6.7in
\centerline{\epsfbox{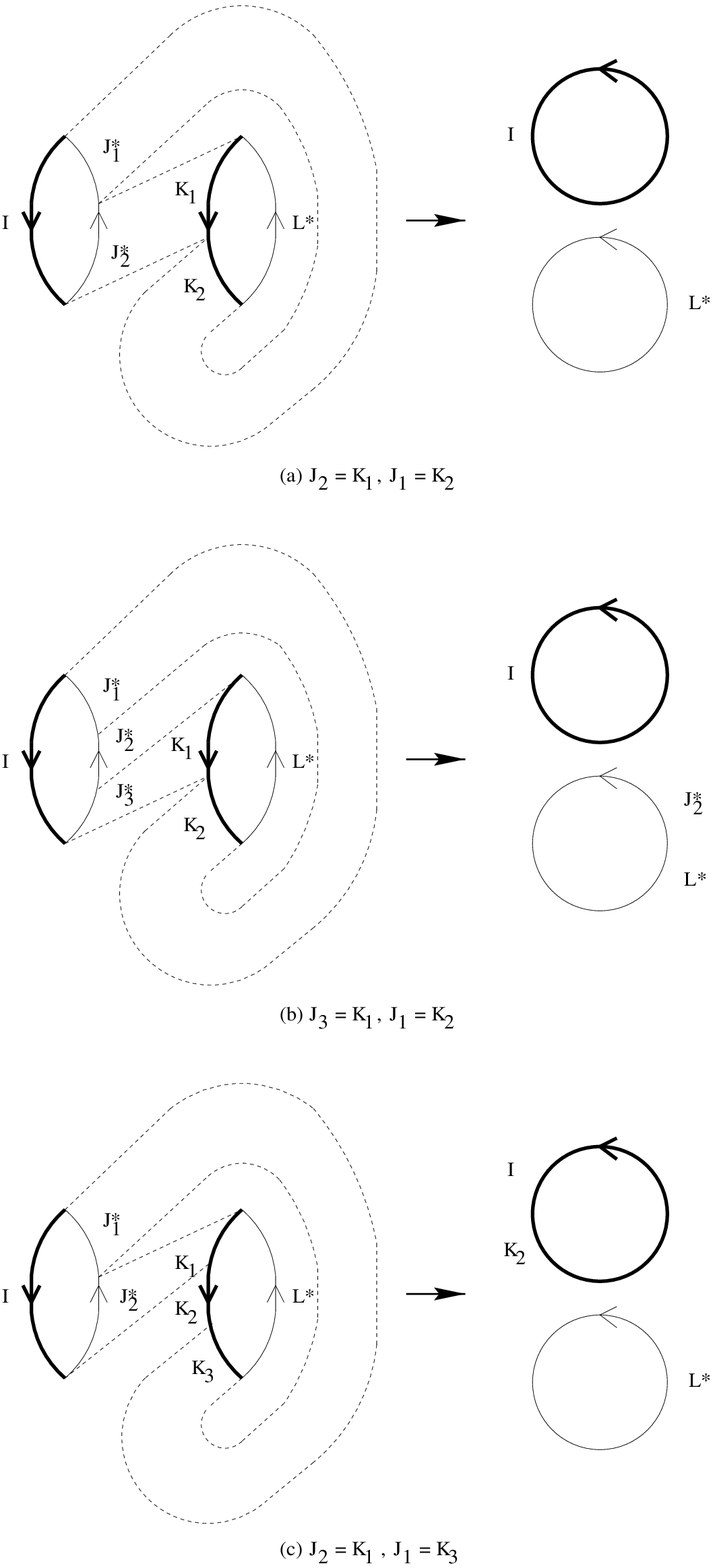}}
\caption{\em Diagrammatic representations of the tenth to twelfth terms on the R.H.S. of 
Eq.(\ref{5.5}).}
\label{f5.4}
\end{figure}

\begin{figure}[ht]
\epsfxsize=3.5in
\centerline{\epsfbox{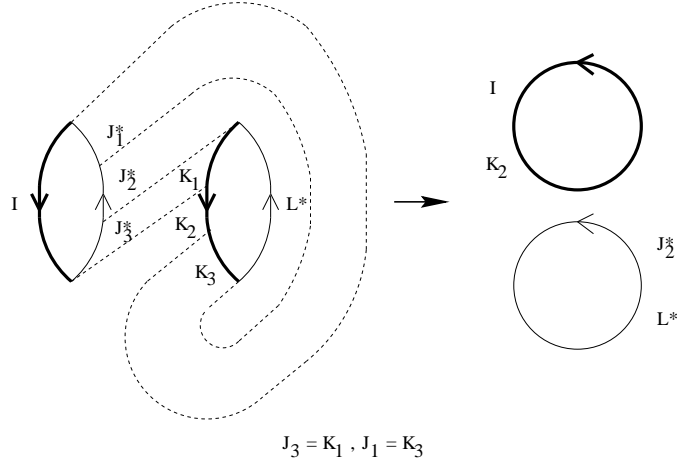}}
\caption{\em Diagrammatic representations of the thirteenth term on the R.H.S. of 
Eq.(\ref{5.5}).}
\label{f5.4.1}
\end{figure}

\begin{figure}
\epsfxsize=4in
\epsfysize=6.5in
\centerline{\epsfbox{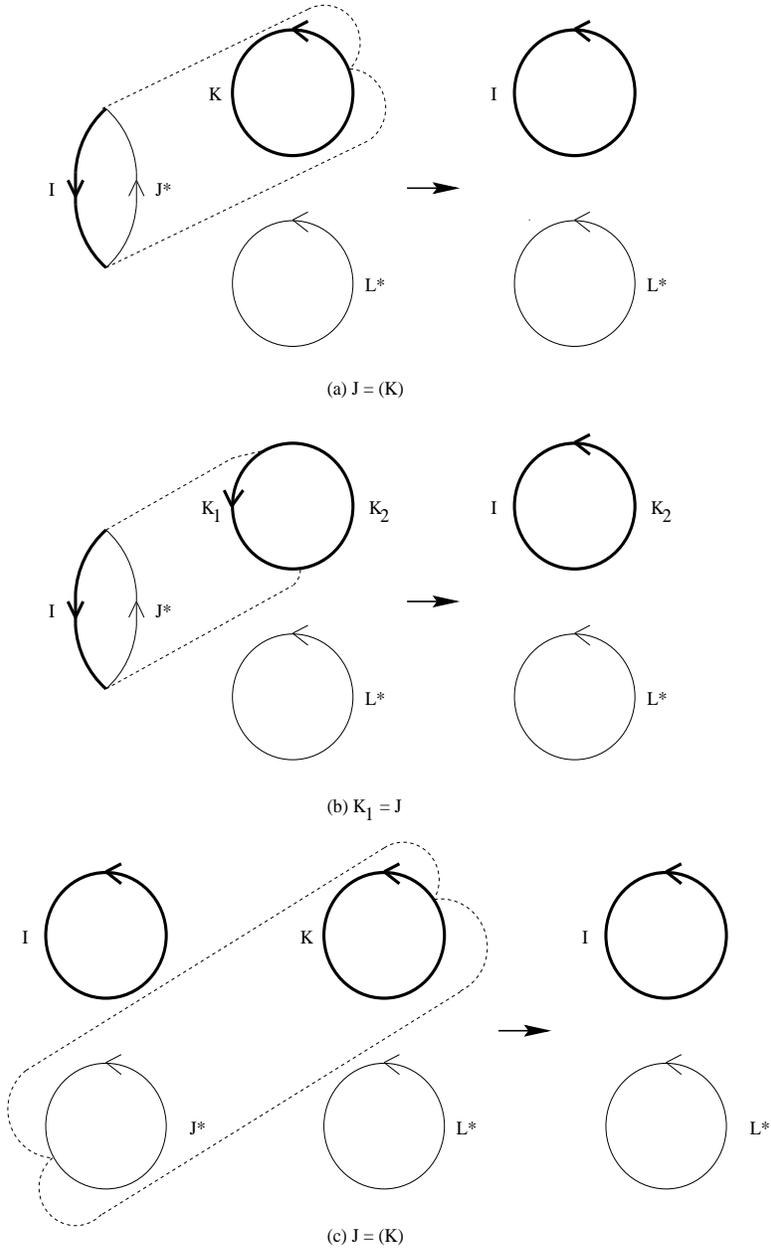}}
\caption{\em Diagrammatic representations of the R.H.S. of Eqs.(\ref{5.6}) and (\ref{5.7}).  Only the first two
terms and the first term are depicted in these two equations respectively.}
\label{f5.5}
\end{figure}

Let us explore the structure of the heterix algebra.  Fist of all,
 we see from Eq.(\ref{5.6}) that $\tsalt$ is a proper ideal of the cyclix 
algebra. Next, consider the subspace  ${\cal H}$  spanned by all vectors of the
forms $\g^I_I$ and $\f^{(I)}_{(I)}$, where $I$ is an arbitrary finite integer sequence of integers between 1 and
$\Lambda$ inclusive.  It turns out that ${\cal H}$ is a Cartan
subalgebra \footnote{ We define, following Ref.~\cite{humphreys},  a Cartan
subalgebra to be a nilpotent subalgebra which is its own normalizer.}
of the heterix algebra.  (The proof can be
found in Appendix~\ref{a8}.)   In addition, we have the following special cases of Eq.(\ref{5.6}) and (\ref{5.7}):
\begin{equation}
   \left[ \g^I_I, \f^{(K)}_{(L)} \right] = \left( \delta^K_{(I)}
   + \sum_{K_1 K_2 = (K)} \delta^{K_1}_I - \delta^I_{(L)} - 
   \sum_{L_1 L_2 = (L)} \delta^I_{L_1} \right) \f^{(K)}_{(L)}
\label{5.8}
\end{equation}
and
\begin{equation}
   \left[ \f^{(I)}_{(I)}, \f^{(K)}_{(L)} \right] = \left(
   \delta^K_{(I)} - \delta^I_{(L)} \right) \f^{(K)}_{(L)}.
\label{5.9}
\end{equation}
Therefore, every $\f^{(K)}_{(L)}$ is a root vector of the cyclix algebra with respect to the subalgebra ${\cal H}$.

Suppose we define  some operators as follows:
\begin{eqnarray}
   l^I_J & \equiv & \g^I_J - \sum_{i=1}^{\Lambda} \g^{iI}_{iJ} - \f^{(I)}_{(J)} \mbox{; and} \nonumber \\
   r^I_J & \equiv & \g^I_J - \sum_{j=1}^{\Lambda} \g^{Ij}_{Jj} - \f^{(I)}_{(J)}.
\label{5.10}
\end{eqnarray}
The actions of these operators on ${\cal T}$ are:
\begin{eqnarray}
   l^I_J s^K & = & \delta^K_J s^I + \sum_{K_1 K_2 = K} \delta^{K_1}_J s^{I K_2}; \nonumber \\
   l^I_J \P^{(K)} & = & 0; \nonumber \\
   r^I_J s^K & = & \delta^K_J s^I + \sum_{K_1 K_2 = K} \delta^{K_2}_J s^{K_1 I} \mbox{; and} \nonumber \\
   r^I_J \P^{(K)} & = & 0.
\end{eqnarray}
Hence they act only on the open string states:
 $l^I_J$ and $r^I_J$ are precisely the operators  defined in a
 previous paper \cite{opstal} on open strings.  From 
the results of that paper, we deduce that the algebra consisting of finite linear combinations of $l^I_J$'s, 
$\hat{L}'_{\Lambda}$, is a subalgebra of the heterix algebra.  Similarly, the algebra consisting of finite linear
combinations of $r^I_J$'s, $\hat{R}'_{\Lambda}$, is also a subalgebra of the cyclix algebra.  

Now define
\begin{eqnarray}
   f^I_J & \equiv & \g^I_J - \sum_{i=1}^{\Lambda} \g^{iI}_{iJ} - \sum_{j=1}^{\Lambda} \g^{Ij}_{Jj}
		    \sum_{i,j=1}^{\Lambda} \g^{iIj}_{iJj} - \f^{(I)}_{(J)} + \sum_{i=1}^{\Lambda} \f^{(iI)}_{(iJ)}
\nonumber \\
   & = & l^I_J - \sum_{j=1}^{\Lambda} l^{Ij}_{Jj} \nonumber \\
   & = & r^I_J - \sum_{i=1}^{\Lambda} r^{iI}_{iJ}.
\label{5.13}
\end{eqnarray}
Its action on ${\cal T}$ reads
\begin{eqnarray}
   f^I_J s^K & = & \delta^K_J s^I \mbox{; and} \nonumber \\
   f^I_J \P^{(K)} & = & 0.
\label{5.11}
\end{eqnarray}
Hence this $f^I_J$ also act only on open strings; it too was introduced in Ref.~\cite{opstal}.  The algebra 
consisting of finite linear combinations of $f^I_J$'s, $F'_{\Lambda}$, is a subalgebra of $\hat{L}'_{\Lambda}$, 
$\hat{R}'_{\Lambda}$ and the heterix algebra.  We deduce from Eqs.(\ref{5.3}), (\ref{5.1}), (\ref{8.1}) and 
(\ref{5.11}) that
\begin{eqnarray}
   \lefteqn{ \left[ \g^I_J, f^K_L \right] = \delta^K_J f^I_L + \sum_{K_1 K_2 = K} \delta^{K_1}_J f^{I K_2}_L
   + \sum_{K_1 K_2 = K} \delta^{K_2}_J f^{K_1 I}_L } \nonumber \\
   & & + \sum_{K_1 K_2 K_3 = K} \delta^{K_2}_J f^{K_1 I K_3}_L 
   - \delta^I_L f^K_J - \sum_{L_1 L_2 = L} \delta^I_{L_1} f^K_{J L_2} \nonumber \\ 
   & & - \sum_{L_1 L_2 = L} \delta^I_{L_2} f^K_{L_1 J} - \sum_{L_1 L_2 L_3 = L} \delta^I_{L_2} f^K_{L_1 J L_3}
   \mbox{; and} \nonumber \\
   \lefteqn{ \left[ \f^{(I)}_{(J)}, f^K_L \right] = 0. }
\end{eqnarray}
In particular,
\begin{eqnarray}
   \left[ \g^I_I, f^K_L \right] & = & \left( \delta^K_I + \sum_{K_1 K_2 = K} \delta^{K_1}_I 
   + \sum_{K_1 K_2 = K} \delta^{K_2}_I + \sum_{K_1 K_2 K_3 = K} \delta^{K_2}_I \right. \nonumber \\
   & & \left. - \delta^I_L - \sum_{L_1 L_2 = L} \delta^I_{L_1} - \sum_{L_1 L_2 = L} \delta^I_{L_2}
   - \sum_{L_1 L_2 L_3 = L} \delta^I_{L_2} \right) f^K_L \mbox{; and} \nonumber \\
   \left[ \f^{(I)}_{(I)}, f^K_L \right] & = & 0.
\label{5.12}
\end{eqnarray}
Thus every $f^K_L$ is also a root vector with respect to the Cartan subalgebra ${\cal H}$.  In fact, a root vector
with respect to ${\cal H}$ is either of the form $\tilde{f}^{(K)}_{(L)}$ or $f^K_L$.  The proof of this assertion 
can be found in Appendix~\ref{a5}.  Hence every root space is one-dimensional.

The reader can refer to Ref.~\cite{opstal} for a more detailed discussion of the Lie algebraic relations
among the $f$'s, $l$'s, $r$'s and $\g$'s.  Also shown in detail in the same article are various diagrammatic
representations of $f$'s, $l$'s, $r$'s, and the Lie algebraic relations pertaining to them.

Now let us consider various quotient algebras from the heterix algebra.  Since $\tsalt$ is a proper ideal, we
can make the set of all cosets $\g^I_J + \tsalt$ into a quotient algebra.  This quotient algebra is nothing but
the centrix algebra $\hat{\Sigma}_{\Lambda}$ \cite{opstal} defined previosly.  The union of $\hat{L}'_{\Lambda}$ 
and $\hat{R}'_{\Lambda}$, the multix algebra $\hat{M}'_{\Lambda}$ \cite{opstal}, is another proper ideal, thus we 
can make the set of all cosets $\g^I_J + \hat{M}'_{\Lambda}$ and $\f^{(I)}_{(J)} + \hat{M}'_{\Lambda}$ a quotient 
algebra as well.  Note that every operator in $\hat{M}'_{\Lambda}$ acts on ${\cal T}_c$ to produce 0.  We 
conjecture that ${\cal T}_c$ provides a faithful representation for the quotient algebra of all cosets 
$\g^I_J + \hat{M}'_{\Lambda}$ and $\tilde{f}^{(I)}_{(J)} + \hat{M}'_{\Lambda}$.  As we will see in the next 
section, the corresponding $\Lambda = 1$ quotient algebra is precisely the Virasoro algebra without any central 
extension.  Therefore the quotient algebra with $\Lambda > 1$ can be regarded as a generalization of the Virasoro 
algebra.

The Lie bracket of the subalgebra $\tsalt$ is given by Eq.(\ref{5.7}).  If we rescale each vector by defining
\begin{equation}
   f^{(I)}_{(J)} \equiv \frac{\f^{(I)}_{(J)}}{\sqrt{\delta^I_{(I)}\delta^J_{(J)}}},
\end{equation}
then the Lie bracket between two $f^{(I)}_{(J)}$'s is similar to that in Eq.(\ref{5.7}),
except that the delta function there with the generic form $\delta^I_{(J)}$ should be 
replaced with another delta function such that this new delta function yields simply 1 if 
$(I) = (J)$, and remains 0 otherwise.  Then there is a one-to-one correspondence 
between each  $f^{(I)}_{(J)}$ and each complex matrix $(a_{ij})_{i,j} \in Z_+$ such that
all but one $a_{ij}$ is nonzero, and that nonzero entry is 1.  Thus the Lie algebras $\tsalt$,
for any $\Lambda$ are all isomorphic to $\gl$, the inductive limit of
the general linear algebras.  $\gl$ and the Kac-Moody algebra associated with it were thoroughly studied, and
their properties can be found in, e.g., Kac's work \cite{kac}. 

\section{The special case $\Lambda = 1$ }
\label{s6}

Let us consider the cyclix algebra for the simplest special case
$\Lambda = 1$
: this is useful as it will connect our Lie algebra with more familiar
algebras such as the Virasoro algebra.     Then all the sequences 
are repetitions of the number $1$ some numbers of times.  We can simplify the 
notations and write $\P^{(K)}$ as $\P^{(\#(K))}$, $s^K$ as $s^{\#(K)}$, $\g^I_J$ as 
$\g^{\#(I)}_{\#(J)}$ and $\f^{(I)}_{(J)}$ as $\f^{(\#(I))}_{(\#(J))}$.  We 
can deduce from Eqs.(\ref{5.3}), (\ref{5.1}) and (\ref{8.1}) that the actions of $\f^{(a)}_{(b)}$ and $\g^a_b$, 
where $a$ and $b$ are the number of integers in the various 
sequences, on $\P^{(c)}$ and $s^c$, where $c$ is also a positive integer, is given by
\begin{eqnarray}
   \f^{(a)}_{(b)} \P^{(c)} & = & c \delta^c_b \P^{(a)}; 
\label{6.1.1} \\
   \f^{(a)}_{(b)} s^c & = & 0; 
\label{6.1.2} \\
   \g^a_b \P^{(c)} & = & c \theta (b \leq c) \P^{(a+c-b)} \mbox{; and}
\label{6.1.3} \\
   \g^a_b s^c & = & (c - b + 1) \theta (b \leq c) s^{a+c-b}.
\label{6.1}
\end{eqnarray}
where $\theta( \mbox{{\em condition}} )$ is 1 if the condition holds, and 
0 otherwise.  The Lie brackets for the $\Lambda = 1$ cyclix algebra are
\begin{eqnarray}
   \left[ \g^a_b, \g^c_d \right] & = & \theta (c \leq b) \left[ 2 \g^{a+c-1}_{b+d-1} + 2 \g^{a+c-2}_{b+d-2} +
   \cdots + 2 \g^{a+1}_{b+d-c+1} \right. \nonumber \\
   & & \left. + (b-c+1) \g^a_{b+d-c} \right] + \theta (2 \leq c \leq b) \left[ \f^{(a+c-2)}_{(b+d-2)} +
   2 \f^{(a+c-3)}_{(b+d-3)} \right. \nonumber \\
   & & \left. + \cdots + (c-1) \f^{(a)}_{(b+d-c)} \right] + \theta (b < c) \left[ 2 \g^{a+c-1}_{b+d-1} +
   2 \g^{a+c-2}_{b+d-2} \right. \nonumber \\
   & & \left. + \cdots + 2 \g^{a+c-b+1}_{d+1} + (c-b+1) \g^{a+c-b}_d \right] + \theta (2 \leq b < c)
   \left[ \f^{(a+c-2)}_{(b+d-2)} \right. \nonumber \\
   & & \left. + \f^{(a+c-3)}_{(b+d-3)} + \cdots + (b-1) \f^{(a+c-d)}_{(b)} \right] \nonumber \\
   & & - (a \leftrightarrow c, b \leftrightarrow d);
\label{6.2}
\end{eqnarray}
\begin{equation}
   \left[ \g^a_b, \f^{(c)}_{(d)} \right] = c \theta (b \leq c) \f^{(a+c-b)}_{(d)} -
   d \theta (a \leq d) \f^{(c)}_{(b+d-a)}
\label{6.2.1}
\end{equation}
and
\begin{equation}
   \left[ \f^{(a)}_{(b)}, \f^{(c)}_{(d)} \right] = b \delta^c_b \f^{(a)}_{(d)} - a \delta^a_d \f^{(c)}_{(b)}.
\end{equation}
These three equations can be deduced from Eqs.(\ref{5.5}), (\ref{5.6}) and (\ref{5.7}) respectively.

The Cartan subalgebra ${\cal H}$ is now spanned by vectors of the forms $\g^a_a$ and $\f^{(a)}_{(a)}$.  A root
vector reads $\f^{(c)}_{(d)}$ or $f^c_d = \g^c_d - 2 \g^{c+1}_{d+1} + \g^{c+2}_{d+2} - \f^{(c)}_{(d)} +
\f^{(c+1)}_{(d+1)}$.  The expression for $f^c_d$ can be deduced from Eq.(\ref{5.13}).  The eigenequations are
\begin{eqnarray}
   \left[ \g^a_a, \f^{(c)}_{(d)} \right] & = & \left( c \theta (a \leq c) - d \theta (a \leq d) \right) 
   \f^{(c)}_{(d)}; \nonumber \\
   \left[ \f^{(a)}_{(a)}, \f^{(c)}_{(d)} \right] & = & a \left( \delta^c_a - \delta^a_d \right) \f^{(c)}_{(d)}; 
\nonumber \\
   \left[ \g^a_a, f^c_d \right] & = & \left( \theta (a \leq c) (c - a + 1) - \theta (a \leq d) (d - a + 1) \right)
   f^c_d \mbox{; and} \\
   \left[ \f^{(a)}_{(a)}, f^c_d \right] & = & 0.
\end{eqnarray}

Let us consider the quotient algebra of all cosets $\g^a_b + \hat{M}'_1$ and $\f^{(a)}_{(b)} + \hat{M}'_1$.  {\em
In the rest of this section and in the accompanying appendices, we will write $\g^a_b + \hat{M}'_1$ and
$\f^{(a)}_{(b)} + \hat{M}'_1$ simply as $g^a_b$ and $\f^{(a)}_{(b)}$ respectively}.  ${\cal T}_c$ still provides a
representation for this quotient algebra --- the actions of $g^a_b$ and $\f^{(a)}_{(b)}$ on ${\cal T}_c$ are still 
described by Eqs.(\ref{6.1.1}) and (\ref{6.1.3}).  However, there is now a relation among $\f$'s and $g$'s:
\begin{equation}
   \f^{(a)}_{(b)} = g^a_b - g^{a+1}_{b+1}.
\label{6.3}
\end{equation}
Hence,
\begin{equation}
   g^a_b = \left\{ \begin{array}{ll}
           - \f^{(a)}_{(b)} - \f^{(a-1)}_{(b-1)} - \cdots - \f^{(a-b+1)}_{(1)} 
           + g^{a-b+1}_1 & \mbox{if $a \geq b$; and} \\
           - \f^{(a)}_{(b)} - \f^{(a-1)}_{(b-1)} - \cdots - \f^{(1)}_{(b-a+1)}
           + g^1_{b-a+1} & \mbox{if $a < b$.}
           \end{array} \right.
\label{a6.2}
\end{equation}
A simplification brought about by setting $\Lambda$ to 1 is that we can write a basis for 
the quotient algebra easily.  Indeed, the set of all $\f^{(a)}_{(b)}$'s, $g^1_b$'s and 
$g^a_1$'s where $a$ and $b$ are arbitrary positive integers form a basis for the 
$\Lambda = 1$ quotient algebra.  The proof of this statement will be given in 
Appendix~\ref{a6}.  The same proof also shows that ${\cal T}_c$ provides not only a representation, but also a 
{\em faithful} representation for this quotient algebra.  From Eq.(\ref{6.2.1}), we deduce that the subspace 
$\tsaltone$ spanned by all the vectors of the form $\f^{(c)}_{(d)}$ form a proper ideal of this quotient algebra.

The set of all $\f^{(a)}_{(a)}$'s, where $a$ is arbitrary, and $g^1_1$ form a basis for a Cartan subalgebra.  The 
proof of this statement will be seen in Appendix~\ref{a7}.  In addition, since $\tsaltone$ is a proper ideal of the 
quotient algebra, the same proof reveals that any root vector with respect to this Cartan subalgebra must be a 
linear combination of a number of $\f^{(a)}_{(b)}$'s.  Therefore all root vectors are of the form 
$\f^{(c)}_{(d)}$'s with $c \neq d$.  The corresponding eigenequations, which can be deduced from Eqs.(\ref{5.8}) 
and (\ref{5.9}), are
\begin{equation}
   \left[ g^1_1, \f^{(c)}_{(d)} \right] = \left( c - d \right) \f^{(c)}_{(d)}
\end{equation}
and
\begin{equation}
   \left[ \f^{(a)}_{(a)}, \f^{(c)}_{(d)} \right] = a \left( \delta^c_a - \delta^a_d \right) \f^{(c)}_{(d)}.
\end{equation}

As $\tsaltone$ is a proper ideal of this $\Lambda=1$ quotient algebra, we can form yet another quotient algebra of 
cosets of the form $v + \tsaltone$ where $v$ is an arbitrary vector of the previous quotient algebra.  This new
quotient algebra is spanned by the cosets $g^a_1 + \tsalt$ and $g^1_b + \tsaltone$, where $a$ and $b$ run over all 
positive integers.  It is a straightforward matter to show that the following Lie brackets are true:
\begin{eqnarray}
   \left[ g^a_1 + \tsaltone, g^c_1 + \tsaltone \right] & = &
   (c - a) \left( g^{a+c-1}_1 + \tsaltone \right) ; \nonumber \\
   \left[ g^a_1 + \tsaltone, g^1_d + \tsaltone \right] & = &
   \left\{ \begin{array}{ll}
   (2 - a - d) \left( g^{a-d+1}_1 + \tsaltone \right) & \mbox{if $d \leq a$, or} \\
   (2 - a - d) \left( g^1_{d-a+1} + \tsaltone \right) & \mbox{if $a \leq d$; and}
   \end{array} \right. \nonumber \\
   \left[ g^1_b + \tsaltone, g^1_d + \tsaltone \right] & = &
   (b - d) \left( g^1_{b+d-1} + \tsaltone \right).
\label{6.4}
\end{eqnarray}
Let us define
\begin{equation}
   L_a = \left\{ 
   \begin{array}{ll}
      - g^{a+1}_1 + \tsaltone & \mbox{if $a \geq 0$, and} \\
      - g^1_{1-a} + \tsaltone & \mbox{if $a \leq 0$.}
   \end{array} \right.
\label{6.5}
\end{equation}
Note that the $L$ here is {\em not} an integer sequence.  Then Eq.(\ref{6.4}) 
becomes
\begin{equation}
   \left[ L_a, L_b \right] = (a - b) L_{a+b}.
\end{equation}
The reader can notice at once that this is the Lie algebraic relation of the Virasoro algebra without any central 
element \cite{7}.  Consequently, the $\M = 1$ quotient algebra of all cosets $\g^a_b$ and 
$\f^{(a)}_{(b)}$ can be regarded as an extension of the Virasoro algebra by an algebra isomorphic to 
$gl_{+\infty}$.  The quotient algebras $\g^I_J + \hat{M}'_{\Lambda} + \tsalt$ for $\Lambda > 1$ can 
then be regarded as generalizations of the Virasoro algebra, as stated in the previous section.

\section{ Canonical Realization }
\label{s2}

We have presented the operators acting on strings somewhat abstractly in
the last section. Although the diagrammatic descriptions give the
definitions of the operators a natural motivation, it would be better
to have a more direct derivation of these relations in terms of more
familiar operators such as those satisfying Canonical Commutation Relations.

Let us define some bosonic operators which satisfy the Canonical Commutation Relations:
\begin{eqnarray}
   \lbrack a^{\mu_1}_{\mu_2} (k_1), a^{\dagger\mu_3}_{\mu_4} (k_2) \rbrack & = & 
   \delta_{k_1 k_2} \delta^{\mu_1}_{\mu_4} \delta^{\mu_3}_{\mu_2}; \nonumber \\
   \lbrack a^{\mu_1}_{\mu_2} (k_1), a^{\mu_3}_{\mu_4} (k_2) \rbrack & = & 0 \mbox{; and}
\nonumber \\
   \lbrack a^{\dagger\mu_1}_{\mu_2} (k_1), a^{\dagger\mu_3}_{\mu_4} (k_2) \rbrack & = & 0; 
\label{1.0.0.1}
\end{eqnarray} 
These operators create and annihilate `gluons', the most fundamental entities in our theory.
The name  `gluons' comes from the application our theory to regularized QCD, in which case these operators act on
gluons.  The indices \m{\mu,\nu} and \m{\upsilon = 1, 2, \ldots, N} label a quantum number we will call `color'.
In this context,  $\Lambda$ is the possible number of
distinct quantum states of a gluon excluding color.  These quantum states are denoted by the numbers 1, 2,
\ldots, $\Lambda$.

We identify the closed string states with a ring of gluons:
\begin{equation}
   \Psi^{(K)} |0\rangle \equiv N^{-c/2} a^{\dagger\upsilon_2}_{\upsilon_1} (k_1) 
   a^{\dagger\upsilon_3}_{\upsilon_2} (k_2) \cdots a^{\dagger\upsilon_1}_{\upsilon_c} (k_c)         
  |0\rangle ,
\label{1.0.2}
\end{equation}
where $c = \#(K)$.  (The summation convention for color indices is implicit in the above and 
following expressions.)  A factor of $N^{-c/2}$ is inserted so that the 
state has a finite norm in the large-$N$ limit.  (Now the beads and connecting lines in Fig.~\ref{f1}(c) 
acquire some extra meanings --- a solid bead represents a creation operator.  Two beads connected by a line 
represents that the two corresponding creation operators share a common color index.  This index is being summed
over.) 
 
We will now look at the operators which can map the above single
glueball states to linear combinations of such states. 
In general they could also create  states with several glueballs but their
amplitudes are suppressed by terms of order at least \m{1\over N}:
this fact is rather well known \cite{th79} and is a reflection in the Hamiltonian
formalism of the planarity of Feynman diagrams in perturbation theory \cite{thooft}.  
As an alternative to reading the relevant papers cited above, the reader can also consult and adapt an argument
in a paper by one of us \cite{lee}, where an operator in
the large-$N$ limit is shown to propagate a single meson state to a linear combination of single
meson states only.

There are two classes of such operators:
\begin{eqnarray}
   \f^{(I)}_{(J)} & \equiv & N^{-(a+b)/2} a^{\dagger\mu_2}_{\mu_1} (i_1)
   a^{\dagger\mu_3}_{\mu_2} (i_2) \cdots a^{\dagger\mu_1}_{\mu_a} (i_a) 
   \nonumber \\
   & & a^{\nu_b}_{\nu_1} (j_b) a^{\nu_{b-1}}_{\nu_b} (j_{b-1}) \cdots a^{\nu_1}_{\nu_2} (j_1).
\label{1.3.1}
\end{eqnarray}
and 
\begin{eqnarray}
   \g^I_J & \equiv & N^{-(a+b-2)/2} a^{\dagger\mu_2}_{\mu_1} (i_1)
   a^{\dagger\mu_3}_{\mu_2} (i_2) \cdots a^{\dagger\nu_b}_{\mu_a} (i_a) 
   \nonumber \\
   & & a^{\nu_{b-1}}_{\nu_b} (j_b) a^{\nu_{b-2}}_{\nu_{b-1}} (j_{b-1}) \cdots
   a^{\mu_1}_{\nu_1} (j_1).
\label{1.3}
\end{eqnarray}
These operators can act on a gluon segment of a glueball.  (Now Figs.~\ref{f5.2} and \ref{f5.1} can be regarded as 
illustrations of these operators.  The solid beads represent creation operators, whereas the hollow beads represent
annihilation operators.  A connecting line, regardless of thickness, symbolizes the fact that
the two corresponding operators share a common color index, and this index is being summed over.)

In the large-$N$ limit, such operators propagate single glueball states to linear combinations of single glueball
states:
\[ \f^{(I)}_{(J)} \P^{(K)} = \delta^K_{(J)} \P^{(I)} \]
and
\[ \g^I_J \P^{(K)} = \delta^K_{(J)} \P^{(I)} + \sum_{K_1 K_2 = (K)} 
   \delta^{K_1}_J \P^{(I K_2)}. \]
These are just the formulae (Eqs.(\ref{5.3}) and (\ref{5.1})) we had in the previous section. 
Thus we have a representation of the Lie algebra of the last section on glueball states.
 
The representation we have obtained is however not faithful: we are
only considering closed string states here.  To emphasize this fact we will denote the operator acting on
closed string states alone by \m{g^I_J}. (We would have a
faithful representation if we included the open string states as
well, but in this paper we will not consider  them.)
Indeed, using Eqs.(\ref{5.1}) and (\ref{5.3}), the reader can show that the following relations hold:
\begin{eqnarray}
   \f^{(I)}_{(J)} & = & g^I_J - \sum_{k=1}^{\Lambda} g^{I k}_{J k} \nonumber \\
   \f^{(I)}_{(J)} & = & g^I_J - \sum_{k=1}^{\Lambda} g^{k I}_{k J} \nonumber \\
   \f^{(I)}_{(J)} & = & g^{I_2 I_1}_J - \sum_{k=1}^{\Lambda}
   g^{I_2 I_1 k}_{J k} \;\mbox{if $I_1 I_2 = I$;} \nonumber \\
   \f^{(I)}_{(J)} & = & g^{I_2 I_1}_J - \sum_{k=1}^{\Lambda}
   g^{k I_2 I_1}_{k J} \;\mbox{if $I_1 I_2 = I$;} \nonumber \\
   \f^{(I)}_{(J)} & = & g^I_{J_2 J_1} - \sum_{k=1}^{\Lambda}
   g^{I k}_{J_2 J_1 k} \;\mbox{if $J_1 J_2 = J$;} \nonumber \\
   \f^{(I)}_{(J)} & = & g^I_{J_2 J_1} - \sum_{k=1}^{\Lambda}
   g^{k I}_{k J_2 J_1} \;\mbox{if $J_1 J_2 = J$;} \nonumber \\
   \f^{(I)}_{(J)} & = & g^{I_2 I_1}_{J_2 J_1} - \sum_{k=1}^{\Lambda}  
   g^{I_2 I_1 k}_{J_2 J_1 k} \;\mbox{if $I_1 I_2 = I$ and $J_1 J_2 = J$; and} \nonumber \\
   \f^{(I)}_{(J)} & = & g^{I_2 I_1}_{J_2 J_1} - \sum_{k=1}^{\Lambda}  
   g^{k I_2 I_1}_{k J_2 J_1} \;\mbox{if $I_1 I_2 = I$ and $J_1 J_2 = J$}.
\label{5.4}
\end{eqnarray}   
These relations are generalizations of Eq.(\ref{6.3}).

The quotient of the heterix algebra by the kernel of this representation was called the cyclix algebra 
\cite{leerajlett} $\hat{C}_{\Lambda}$.  As have been stated in a previous section, we conjecture that
$\hat{C}_{\Lambda} = \hat{\Gamma}_{\Lambda} / \hat{M}'_{\Lambda}$.  The results in Section~\ref{s6} shows
that this conjecture is true for $\Lambda = 1$.

\section{Application to Gauge Field Theory}
\label{s9}

We are going to apply the cyclix algebra to
physical models which can be formulated as matrix models in the large-$N$ limit.  In this section, we will
concentrate on gauge field theories in the large-$N$ limit.  We will find that we can express the physical 
observables of these theories in terms of the elements of the cyclix algebra.  
(Indeed, we will see in a future paper \cite{sustal} that closed superstring models and M-theory can be formulated 
in terms of the supersymmetric generalization of the cyclix algebra in a similar manner.)  However, they will look 
complicated.  In the next section, we will construct multi-matrix models {\em integrable} in the large-$N$ limit
associated with integrable quantum spin chain models satisfying the periodic boundary condition.  Studying the
ways these multi-matrix models are solved can shed light on how the more complicated gauge field theories are
solved.

In this section we will let the regulator \m{\Lambda\to \infty}, so
that the momentum indices will take an infinite number of values. In
order that the previous discussions of our algebra apply directly here, we will
need to regularize the field theory such that the momentum variables
can take only a finite number \m{\Lambda} of distinct values. But this
is mostly a  technicality, since  we will only talk of field theories without
divergences for which the limit \m{\Lambda\to \infty} should
exist: dimensional reductions of gauge theory to 1+1 dimesnions.
 The deeper problem of describing renormalizable field theories
this way is being studied.

The first gauge field model is a (1+1)-dimensional SU($N$) gauge theory coupled to bosonic 
matter in the adjoint representation.  This model has been studied by Dalley and Klebanov \cite{DaKl93a}
in detail and so we will only give a brief account here.  Let $g$ be the strong coupling constant, $\alpha$ and 
$\beta$ be ordinary spacetime indices, $A_{\alpha}$ be a gauge potential and $\phi$ be a scalar field in the 
adjoint representation of the gauge group $U(N)$.  Both $\phi$ and $A_{\alpha}$ are  
$N\times N$ Hermitian matrix fields. (We can regard both \m{A_\alpha}
and \m{\phi} as describing the gluons of a Yang-Mills theory in three
dimensions, dimensionally reduced to 1+1 dimensions.) Let the covariant derivative $D_{\alpha} \phi = 
\partial_{\alpha}\phi + {\rm i} \lbrack A_{\alpha},\phi \rbrack$ and the Yang-Mills field 
$F_{\alpha\beta}= \partial_{\alpha} A_{\beta}  - \partial_{\beta} A_{\alpha} + {\rm i}[A_{\alpha}, 
A_{\beta}]$.  Lastly, let $m$ be the mass of a boson and $g$ be the gauge field coupling 
strength.  Then the Minkowski space action is 
\begin{equation}
   S=\int d^2 x\,{\rm Tr} \left[ \frac{1}{2} D_{\alpha}\phi D^{\alpha}\phi -
   \frac{1}{2} m^2 \phi^2 - \frac{1}{4g^2} F_{\alpha\beta} F^{\alpha\beta} 
\right].
\end{equation}
We then transform the Minkowski coordinates to light-cone coordinates, and choose the 
light-cone gauge $A_- = (A_0 - A_3) / \sqrt{2} = 0$.  Then $A_+$ is not dynamical and can be 
eliminated by the constrained equations.  We canonically quantize the adjoint matter field:
\begin{equation}
   \phi^{\mu}_{\nu} (x^+ = 0) =\frac{1}{\sqrt{2\pi}} \int_0^{\infty} \frac{dk^+}{\sqrt{2k^+}}   
   \left[ a^{\mu}_{\nu} (k^+) {\rm e}^{-{\rm i}k^+ x^-} + a^{\dagger\mu}_{\nu} (k^+) 
   {\rm e}^{{\rm i}k^+ x^-} \right]
\label{9.0.1}
\end{equation}
where $\mu$ and $\nu$ are color indices.  The creation and annihilation operators satisfy the Canonical Commutation 
Relation Eq.(\ref{1.0.0.1}).  One of the constraints dictate that the Hilbert space on which the creation and 
annihilation operators act be spanned by color singlet states:
\begin{equation}
   N^{-c/2} {\rm Tr} \left[ a^{\dagger} (k_1^+) \cdots a^{\dagger} (k_c^+) \right] |0 \rangle.
\label{9.1}
\end{equation}
Eq.(\ref{9.1}) is exactly in the form of a single glueball state $\Psi^{(K)} |0\rangle$ 
defined in Eq.(\ref{1.0.2}).  Thus the creation operator can be interpreted as a creation 
operator for a gluon.  Furthermore, We can obtain the light-cone momentum and energy in 
terms of the elements of the cyclix algebra:
\begin{eqnarray}
   P^+ & = & \int_0^{\infty} dk \; k g^k_k \mbox{; and} 
\label{9.2} \\
   P^- & = & \frac{1}{2} m^2 \int_0^{\infty} \frac{dk}{k} g^k_k 
   + \frac{g^2 N}{4\pi} \int_0^{\infty} \frac{dk}{k} C g^k_k \nonumber \\
   & & + \frac{g^2 N}{8\pi} \int_0^{\infty} \frac{dk_1 dk_2 dk_3 dk_4}
   {\sqrt{k_1 k_2 k_3 k_4}}  
   \left\{ A \delta (k_1 + k_2 - k_3 - k_4) g^{k_1 k_2}_{k_3 k_4} \right. \nonumber \\  
   & & + B \delta (k_1 - k_2 - k_3 - k_4) g^{k_1}_{k_2 k_3 k_4} \nonumber \\
   & & + \left. B \delta (k_1 + k_2 + k_3 - k_4) g^{k_1 k_2 k_3}_{k_4} \right\}
\label{9.3}
\end{eqnarray}     
where 
\begin{eqnarray*}
   A & = & \frac{(k_2 - k_1)(k_4 - k_3)}{(k_1 + k_2)^2} - \frac{(k_3 + k_1)(k_4 + k_2)}{(k_4 - k_2)^2}; \\
   B & = & \frac{(k_1 + k_4)(k_3 - k_2)}{(k_3 + k_2)^2} + \frac{(k_3 + k_4)(k_1 - k_2)}{(k_1 + k_2)^2} \mbox{; and}
   \\
   C & = & \int_0^k dp \frac{(k+p)^2}{p(k-p)^2}.
\end{eqnarray*}

This adjoint matter model can be used to study glueball spectrum
\cite{AnDa}.  Consider next a 
(3+1)-dimensional QCD model with only gluons and no quarks.  If we apply the dimensional 
reduction  to it, we will obtain an effective Yang-Mills model in 1+1
dimensions with 2 adjoint 
matter fields $\phi_1$ and $\phi_2$, which are constant multiples of transverse gluon fields.  Rewrite the adjoint
matter fields in the helicity basis
\begin{equation}
   \phi_{\pm} = \frac{\phi_1 \pm {\rm i} \phi_2}{\sqrt{2}},
\end{equation}
and canonically quantize $\phi_{\pm}$:
\begin{equation}
   \phi^{\mu}_{\pm\nu} (x^+ = 0) =\frac{1}{\sqrt{2\pi}} \int_0^{\infty} \frac{dk^+}{\sqrt{2k^+}}   
   \left[ a^{\mu}_{\nu} (k^+, \pm) {\rm e}^{-{\rm i}k^+ x^-} + a^{\dagger\mu}_{\nu} (k^+, \pm) 
   {\rm e}^{{\rm i}k^+ x^-} \right].
\end{equation}
Then the light-front momentum and energy are
\begin{eqnarray}
   P^+ & = & \int_0^{\infty} dk \; k \sum_{j = +, -} g^{k, j}_{k, j} \mbox{; and} \nonumber \\
   P^- & = & \frac{m_{\mbox{ind}}^2}{2} \int_0^{\infty} \frac{dk}{k} g^{k, +}_{k, +} 
   + \frac{g^2}{8 \pi} \int_0^{\infty} \frac{dk_1 dk_2 dk_3 dk_4}{\sqrt{k_1 k_2 k_3 k_4}}
   \delta(k_1 + k_2 - k_3 - k_4) \cdot \nonumber \\
   & & \left\{ (D \lbrack k_1, k_2, k_3 \rbrack - 1) g^{k_3, +; k_4, -}_{k_1, -; k_2, +} +
   (-E \lbrack k_1, k_2, k_3 \rbrack - 1) g^{k_3, +; k_4, +}_{k_1, +; k_2, +} \right. \nonumber \\
   & & \left. + (D \lbrack k_1, k_2, k_3 \rbrack - E \lbrack k_1, k_2, k_3 \rbrack + 2) 
   g^{k_3, +; k_4, -}_{k_1, +; k_2, -}
   \right\} \nonumber \\ 
   & & + \frac{g^2}{8 \pi} \int_0^{\infty} \frac{dk_1 dk_2 dk_3 dk_4}{\sqrt{k_1 k_2 k_3 k_4}}
   \delta(k_1 + k_2 + k_3 - k_4) \cdot \nonumber \\
   & & \left\{ (F \lbrack k_1, k_2, k_3 \rbrack + 1) (g^{k_1, +; k_2, -; k_3, +}_{k_4, +} +
   g^{k_3, +; k_2, -; k_1, +}_{k_4, +} + g^{k_4, +}_{k_1, +; k_2, -; k_3, +} \right. \nonumber \\
   & & + g^{k_4, +}_{k_3, +; k_2, -; k_1, +}) \nonumber \\
   & & + (F \lbrack k_1, k_2, k_3 \rbrack - 1) (g^{k_3, -; k_2, -; k_1, +}_{k_4, -} +
   g^{k_1, +; k_2, -; k_3, -}_{k_4, -} + g^{k_4, +}_{k_3, +; k_2, +; k_1, -} \nonumber \\
   & & \left. + g^{k_4, +}_{k_1, -; k_2, +; k_3, +}) \right\} \nonumber \\
   & & + (\mbox{positive helicity} \leftrightarrow \mbox{negative helicity})
\end{eqnarray}
where
\begin{eqnarray*}
   E \lbrack x_1, x_2, x_3 \rbrack & = & \frac{(x_1 + x_3)(x_1 + 2 x_2 - x_3)}{(x_1 - x_3)^2}; \\
   D \lbrack x_1, x_2, x_3 \rbrack & = & E \lbrack x_2, -x_3, -x_1 \rbrack \mbox{; and} \\
   F \lbrack x_3, x_2, x_1 \rbrack & = & E \lbrack x_3, x_1, -x_2 \rbrack,
\end{eqnarray*}
and $m_{\mbox{ind}}$ is the induced mass due to normal-ordering the creation and annihilation operators.

The various algebras described in this paper can also be modified to study supersymmetric 
models with fermionic adjoint matter fields.  Not long ago there were studies on $d>2$ 
noncritical superstrings using a matrix model \cite{HaKl}.  The superfields were presented as 
matrices.  Canonical quantization yielded a number of operators satisfying various canonical 
commutation and anticommutation relations.  The Hilbert space on which these operators acted 
were composed of closed strings of bosonic and fermionic creation operators.  The terms in 
the supercharge operators could be written in a way analogous to the gluonic operators, 
except that there were fermionic operators in addition to bosonic operators, which are the 
only operators present in the gluonic operators defined in Eq.(\ref{1.3}).  The various 
algebras described in this paper can be generalized to superalgebras \cite{sustal} to study this 
noncritical superstring model.  A closely related model is a supersymmetric Yang-Mills theory with a fermionic 
matter field \cite{DaKl93a, kutasov}.  Moreover, these superalgebras can also be employed to study 
M-theory, which is being conjectured as a matrix model \cite{susskind}. 

\section{Solvable Matrix And Spin Chain Models}
\label{s7}

Now that we have learnt how to formulate gauge field theories in the large-$N$ limit in the language of the cyclix
algebra, let us turn to integrable periodic quantum spin chain models and the associated, more tractable 
multi-matrix models.  Understanding the properties of these solvable matrix models may shed new light on the
mathematics of the more difficult gauge field theories.

Consider the Hamiltonian of a matrix model in the form of the linear combination $H = \sum_{IJ} h^J_I g^I_J$ where 
$h^J_I \neq 0$ only if $I$ and $J$ have the same number of indices.  (This means that the `parton number' is 
conserved.  Note that the gauge field theories described in the previous section are not of this type.)  Such 
linear combinations form a subalgebra of the cyclix algebra.  Let us call this $\hat{C}^0_{\Lambda}$.

{\em There is an isomorphism between multi-matrix models whose Hamiltonians are in $\hat{C}^0_{\Lambda}$ and 
periodic quantum spin chains.}  Consider a spin chain with $c$ sites.  At any site $p = 1, 2$, \ldots, or $c$
there is a variable $i_p$ called {\em spin} that describes the quantum state of that site, and that can take the 
value 1, 2, \ldots, or $\Lambda$.  We impose the periodic boundary
condition.  We will show that a subspace of these states, namely, the states with zero total momentum, provides a
(non-faithful) representation of the algebra \m{\hat{C}^0_{\Lambda}}. 

A basis of states is given by the set
of states of the form $|k_1 k_2 \ldots k_c \rangle$.  Define the operator $X^i_j(p)$ by its action on these 
collective spin chain states by the formula 
\begin{equation}
   X^i_j(p) |k_1 k_2 \ldots k_c \rangle \equiv \delta^{k_p}_j |k_1 k_2 \ldots k_{p-1} i k_{p+1} \ldots k_c \rangle.
\end{equation}
(Those who are familiar with the theory of the Hubbard model can notice at once that $X^i_j(p)$ is the Hubbard 
operator at the site $p$, which is conventionally written as $X^{ij}_p$ \cite{bablog}.)  Furthermore, we impose the
periodic boundary condition $X^i_j(p + c) = X^i_j(p)$.  It can be
verified that if $I$ and $J$ have the same length $a \leq c$, then
\begin{equation}
   r_c (g^I_J) \equiv \sum_{p=1}^c X^{i_1}_{j_1}(p) X^{i_2}_{j_2}(p+1) \cdots X^{i_a}_{j_a}(p+a-1)
\end{equation}
satisfies the commutation relations of the algebra $\hat{C}^0_{\Lambda}$.  If we further define $r_c (g^I_J) = 0$ 
for $a > c$, we will have a representation $r_c$ of $\hat{C}^0_{\Lambda}$.  The cyclically symmetric states of the
matrix model correspond to the states of the periodic spin chain with
zero total momentum. The representation is not faithful as it is sets
all generators of length \m{a} greater than  the number of spins 
\m{c} to zero; if we take spin chains of all possible  lengths, we will
get a  faithful representation.

We are now ready to associate the matrix model which the Hamiltonian $H = \sum_{IJ} h^J_I g^I_J$ is in
$\hat{C}^0_{\Lambda}$, with the quantum spin chain with the Hamiltonian
\begin{equation}
   H^{\spin} = \sum_{IJ} h^J_I \sum_{p=1}^c X^{i_1}_{j_1}(p) X^{i_2}_{j_2}(p+1) \cdots X^{i_a}_{j_a}(p+a-1).
\end{equation}
Therefore, matrix models conserving the parton number correspond to quantum spin systems with interactions 
involving neighborhoods of spins at the sites $p$, $p+1$, \ldots, and $p+a-1$.

Recall that a convenient way of solving spin chain systems \cite{GoRuSi}
is via the Bethe ansatz.  This ansatz is applicable whenever the spin wave S-matrix satisfies the 
Yang-Baxter equation, in which case we say that the spin chain system is integrable.  Then this integrable spin
chain system yield an integrable multi-matrix model.

Let us look at the simplest example, the quantum Ising spin chain \cite{onsager, kogut}.  
The Hamiltonian $H^{\rm spin}_{\rm Ising}$ of this model is
\beq
   H^{\rm spin}_{\rm Ising}(\tau, \lambda) = \sum_{p=1}^c \tau^z(p) + \lambda \sum_{p=1}^c \tau^x(p) \tau^x(p+1).
\label{7.1}
\eeq
Here $\lambda$ is a constant, and \m{\tau^{x,y,z}_p} are Pauli matrices at site \m{p}.  Two Pauli matrices at different sites (i.e., with different 
subscripts)  commute with each other.  Let us rewrite the Ising spin chain as a two-matrix model.  We set the quantum 
state 1 for a boson in the matrix model to correspond to the spin-up state in the Ising spin chain, and the quantum
state 2 to correspond to the spin-down state.  Since
\begin{equation}
   \tau^z_p = X^1_1(p) - X^2_2(p), \; \tau^x_p + {\rm i} \tau^y_p = 2 X^1_2(p) \; \mbox{and} \;
   \tau^x_p - {\rm i} \tau^y_p = 2 X^2_1(p),
\end{equation}
we obtain the corresponding element in $\hat{C}^0_2$:
\beq
	H^{\rm matrix}_{\rm Ising} = H_0 + \lambda V
\eeq
where
\beq
   H_0 & = & g^1_1 - g^2_2 \mbox{; and} \nonumber \\
     V & = & \lambda \left[ g^{22}_{11}+g^{21}_{12}+g^{12}_{21}+g^{11}_{22} \right].
\label{7.1.1}
\eeq
This is a two-matrix model with the Hamiltonian
\beq
   H^{\rm matrix}_{\rm Ising}&=&{\rm tr} [a^{\dag}(1)a(1)-a^{\dag}(2)a(2)]
   +{\lambda \over N}{\rm tr} \left[ a^{\dag }(2)a^{\dag }(2)a(1)a(1) \right. \nonumber \\
   & & +
   a^{\dag }(2)a^{\dag }(1)a(2)a(1)+a^{\dag }(1)a^{\dag }(2)a(1)a(2) \nonumber \\
   & & \left. + a^{\dag }(1)a^{\dag }(1)a(2)a(2) \right] .
\label{7.1.2}
\eeq
Our results, along with known results of the Ising spin chain \cite{kogut}, give the
spectrum of this matrix model in the large \m{N} limit:
\beq
	E(n_p, c, \lambda) = -2\sum_{p=-c}^c \Bigg(1+2\lambda \cos\big[{2\pi p\over 2c +1} \big] + 
			     \lambda^2\Bigg)^{1/2}n_p
\label{7.2}
\eeq
where \m{c} is any positive integer and \m{n_p = 0} or 1.  Also, we must impose the condition  
\m{\sum_{p=-c}^c n_p p=0} to get cyclically symmetric states.  Let us underscore that {\em the matrix model defined 
by the Hamiltonian $H^{\matrix}_{\rm Ising}$ in Eqs.(\ref{7.1.1}) or (\ref{7.1.2}) is an integrable matrix model in 
the large-$N$ limit}.

Eq.(\ref{7.2}) manifests the self-duality of the Ising model:
\beq
   E(n_p, c, \lambda) = \lambda E(n_p, c, \frac{1}{\lambda}).
\eeq
Indeed, under the operator dual transformation
\beq
   \tilde{\tau}^x(p) & \equiv & \tau^z(p) \tau^z(p+1) \; \mbox{and} \nonumber \\
   \tilde{\tau}^z(p) & \equiv & \tau^x(1) \tau^x(2) \cdots \tau^x(p),
\eeq
the Hamiltonian is changed to
\beq
   H^{\rm spin}_{\rm Ising}(\tilde{\tau}, \lambda) =  \sum_{p=1}^c \tilde{\tau}^x(p) \tilde{\tau}^x(p+1) + 
   \lambda \sum_{p=1}^c \tilde{\tau}^z(p).
\eeq
Since $\tilde{\tau}^x(p)$'s and $\tilde{\tau}^z(p)$'s satisfy the same algebra as $\tau^x(p)$'s and $\tau^z(p)$'s,
\beq
   H^{\rm spin}_{\rm Ising}(\tilde{\tau}, \lambda) = \lambda H^{\rm spin}_{\rm Ising}(\tau, \frac{1}{\lambda}), 
\eeq
i.e., the Ising model is self-dual.

It is possible to understand the solvability of the Ising model in
terms of the  Dolan-Grady conditions \cite{dogr, da} and the Onsager
algebra. 
Let us digress to summarize this method.
Suppose we have a  system whose hamiltonian can be written as 
\beq
	H=H_0+V
\eeq
with  the two terms in the hamiltonian satisfying the Dolan-Grady conditions:
\beq
   \lbrack H_0, \lbrack H_0, \lbrack H_0, V \rbrack \rbrack \rbrack = 16 \lbrack H_0, V \rbrack,
\label{7.3}
\eeq
and 
\beq
   \lbrack V, \lbrack V, \lbrack V, H_0 \rbrack \rbrack \rbrack = 16 \lbrack V, H_0 \rbrack.
\label{7.4}
\eeq
Then we can construct operators satisfying an infinite-dimensional Lie algebra
 \beq  
   \lbrack A_m, A_n \rbrack = 4 G_{m-n}, \; \lbrack G_m, A_n \rbrack = 2 A_{n+m} - 2 A_{n-m},
   \; \mbox{and} \; \lbrack G_m, G_n \rbrack = 0
\eeq
by the following recursion relations:
\beq
   & A_0 = H_0, \; A_1 = V, \; A_{n+1} - A_{n-1} = \frac{1}{2} \lbrack G_1, A_n \rbrack, & \nonumber \\
   & G_1 = \frac{1}{4} \lbrack A_1, A_0 \rbrack, \; \mbox{and} \; G_n = \frac{1}{4} \lbrack A_n, A_0 \rbrack. &
\eeq
This Lie algebra is called the Onsager Lie algebra. It is known to be
 isomorphic to an infinite direct sum of \m{su(2)} algebras. In particular, the
 system will admit an infinite number of conserved quantities \cite{honecker}:
\beq
   Q_m = -\frac{1}{2} \left( A_m + A_{-m} + \lambda A_{m+1} + \lambda A_{-m+1} \right).
\eeq
Thus any such system should be integrable.

In the case of the Ising model we  choose \m{H_0} and \m{V} as
above. The Lie brackets of the \m{g^I_J} then allow us to verify
easily that the first of the Dolan-Grady conditions is satisfied.
This, together with the self-duality of the Ising model, guarantee that the other Dolan-Grady condition
is satisfied also.  (Eq.(\ref{7.4}) could also be verified directly.  The only caveat is that as the operators act 
on closed string states only, we have to treat all $\f$'s as linear combinations of $g$'s in order for 
Eq.(\ref{7.4}) to hold true.)  
Moreover we see that the Onsager algebra is a subalgebra of our
algebra \m{\hat{C}^0_\Lambda}: all the conserved quantities are just
linear combinations of our \m{g^I_J} and \m{\f^I_J}.
This suggests that there may  be other models that are integrable by
this method: we need to identify pairs of elements in our algebra that
satisfy the Dolan-Grady conditions.

We can transcribe other integrable quantum spin chain models with the periodic
boundary condition to integrable matrix models in the large-$N$
limit. Some examples are listed below:
\begin{itemize}
\item{\em a generalization of the Ising model} \cite{honecker}.  The Hamiltonian is
\beq
	H^{\rm spin}_{\rm GI}=\sum_{p=1}^c \tau^z_p + \lambda \sum_{p=1}^c 
	[\tau^x_p \tau^x_{p+1} + v \{ \tau^x_p \tau^y_{p+1}- \tau^y_p \tau^x_{p+1}\}],
\eeq 
where $\lambda$ and $v$ are constants.  The corresponding Hamiltonian in the matrix model is
\beq
	H^{\rm matrix}_{\rm GI}&=&g^1_1-g^2_2+
	\lambda \left[ g^{22}_{11}+(1-2iv)g^{21}_{12}+(1+2iv)g^{12}_{21}+g^{11}_{22} \right].
\eeq
This model satisfies the Dolan-Grady conditions and hence is
integrable using the Onsager subalgebra.

\item{\em the XYZ model} \cite{baxter}.  This is a generalization of the Ising model in another
direction.  It doesnt satisfy the Dolan-Grady condition, but is
integrable by methods using the Yang-Baxter equation. 
 The Hamiltonian is
\beq
	H^{{\rm spin}}_{{\rm XYZ}} = \sum_{p=1}^c \tau^z_p \tau^z_{p+1}-
	\lambda \sum_{p=1}^c \left[ \tau_p^x \tau^x_{p+1} +v \tau^y_p \tau^y_{p+1} \right].
\eeq
The corresponding Hamiltonian in the matrix model is
\beq
	H^{{\rm matrix}}_{{\rm XYZ}} & = & g^{11}_{11}-g^{12}_{12}-g^{21}_{21}+g^{22}_{22}- \nonumber \\
	 & & \lambda \left[ (1-v)(g^{22}_{11}+g^{11}_{22})+(1+v)(g^{21}_{12}+g^{12}_{21}) \right].
\eeq
\item{\em the chiral Potts model} \cite{chPotts}.  This is a model in which the number of quantum states available
for a site is not restricted to 2 but is any finite positive integer.  The Hamiltonian is
\beq
	H^{{\rm spin}}_{{\rm CP}}=\sum_{p=1}^c \sum_{k=1}^{\M -1}[\tilde \alpha_k Q_p^k + 
        \lambda \alpha_k P_p^k P^{\M-k}_{p+1}],
\eeq
where
\begin{eqnarray*}
   & \alpha_k = \frac{{\rm e}^{{\rm i} \phi (\frac{2k}{n} - 1)}}{\sin \frac{\pi k}{n}}, \;
   \tilde{\alpha}_k = \frac{{\rm e}^{{\rm i} \varphi (\frac{2k}{n} - 1)}}{\sin \frac{\pi k}{n}}, \;
   \cos \varphi = \lambda \cos \phi, \; \omega = {\rm e}^{\frac{2 \pi {\rm i}}{n}}, & \nonumber \\
   & \sigma_j = \left( \begin{array}{ccccc}
   			1 & 0 & 0 & \ldots & 0 \\
   			0 & \omega & 0 & \ldots & 0 \\
   			0 & 0 & \omega^2 & \ldots & 0 \\
   			& & \ldots & & \\
   			0 & 0 & 0 & \ldots & \omega^{n-1}
   		     \end{array} \right), &
\end{eqnarray*}
and
\begin{equation}
   \Gamma_j = \left( \begin{array}{cccccc}
   			0 & 0 & 0 & \ldots & 0 & 1 \\
   			1 & 0 & 0 & \ldots & 0 & 0 \\
   			0 & 1 & 0 & \ldots & 0 & 0 \\
   			  &   &   & \ldots &   &   \\
   			0 & 0 & 0 & \ldots & 1 & 0
   		     \end{array} \right).
\end{equation}
This model is exactly solvable by the Yang-Baxter method. 
The Hamiltonian of the associated solvable multi-matrix model is
\beq
   H^{\rm matrix}_{\rm CP} = \sum_{k=1}^{\M-1} \left[ \tilde\alpha_k \sum_{j=1}^{\M} \omega^{k(j-1)} g^j_j +
   \lambda \alpha_k \sum_{j_1, j_2 = 1}^{\M} g^{j_1 + k, j_2 - k}_{j_1,j_2} \right]
\eeq
where $j_1 + k$ should be replaced with $j_1 + k - \M$ if $j_1 + k > \M$ and $j_2 - k$ should be replaced with
$j_2 + \M - k$ if $j_2 - k \leq 0$ in $g^{j_1 + k, j_2 - k}_{j_1, j_2}$ in the above equation.
\end{itemize}

\vskip 1pc
\noindent \Large{\bf \hskip .2pc Acknowledgments}
\vskip 1pc
\noindent 

\normalsize
We thank O. T. Turgut for discussions in an early stage of this
work. S. G. R. thanks the I.H.E.S., where part of this work was done,
 for hospitality. We were  supported in part by funds provided by the U.S.
Department of Energy under grant DE-FG02-91ER40685.

\vskip 1pc
\noindent \Large{\bf \hskip .2pc Appendix}

\normalsize
\appendix

\section{Notation for Multi-Indices}
\label{s1-1}

Much of our work involves manipulating tensors carrying multiple indices.
For the convenience of the reader, we give here a summary of the notations used
in this paper for multi-indices.  We have attempted to match the
notations with our other papers \cite{leerajlett, opstal} and to make
this Appendix self-contained. More details can be found in another paper \cite{opstal}.

We will use lower case Latin letters such as \m{i, j, i_1} and \m{j_2} to
denote indices which are positive integers 1, 2, \ldots, and \m{\Lambda}. Here \m{\Lambda}
itself is a fixed positive integer, denoting the number of degrees of
freedom of gluons. A  \emph{non-empty} sequence of indices \m{i_1i_2i_3\ldots i_a}
 will be denoted  by the corresponding uppercase letter \m{I}. The
length of the  sequence \m{I} will be denoted by \m{\#(I)}. 

A capital letter \m{I} denotes a {\em non-empty} sequence of indices
\m{i_1i_2\cdots i_a}, each taking values from the set \m{\{1,2,\cdots \Lambda\}}.
The length \m{\#(I)} of \m{I} is just the number of elements in the sequence.
Two sequences are equal if they have the same entries.
Concatenation of two sequences will be denoted by 
\[ I J = i_1 i_2\ldots i_a j_1 j_2 \ldots j_b. \]
In particular, 
\beq
	Ij=i_1i_2\ldots i_aj,
\eeq
when only a single index is added at the end. 

The Kronecker symbol has the obvious definition:
\[ \delta^I_J \equiv \left\{ \begin{array}{ll}
   				1 & \mbox{if $I = J$; or} \\
   				0 & \mbox{if $I \neq J$.} \\
   			     \end{array} \right\}. \]
Thus, 
\beq
   \sum_{I_1 I_2 = I} X^{I_1 }Y^{I_2}_K \equiv 
   \sum_{I_1I_2}\delta_I^{I_1I_2}X^{I_1}Y^{I_2}_K,
\label{1-1.1}
\eeq
where $X^{I_1}$ is a function dependent on $I_1$ and $Y^{I_2}_K$ is another function dependent on $I_2$ and $K$,
denotes the sum over all the ways in which a given index \m{I}
can be split into two {\em nonempty} subsequences \m{I_1} and
\m{I_2}. If there is no way to split $I$ as required, then the sum simply yields 0.

We define $(I)$ to be the equivalence 
class of all cyclic permutations of $I$. In fact \m{(I)} can be viewed
as a discrete model for a closed loop, or closed string.
The corresponding Kronecker delta function is defined by the following relation:
\begin{equation}
   \delta^I_{(J)} \equiv \delta^I_J + \sum_{J_1 J_2 = J} \delta^I_{J_2 J_1}. 
\label{1-1.2}
\end{equation}
Eq.(\ref{1-1.2}) means that the delta function returns the number of different 
cyclic permutations of $J$ such that each permuted sequence is identical with 
$I$. {\em Thus \m{\delta^I_{(J)}} can take  any non-negative integer as
its value, not just 0 or 1.}  The reader can verify from this definition that
\begin{equation}
    \delta^I_{(J)} = \delta^J_{(I)} 
\label{1-1.2.1}
\end{equation}
Next, the expression
\[ \sum_{(I)} X^I, \]
where $X^I$ is dependent on the equivalence class $(I)$, i.e., $X^I = X^J$ if
$(I) = (J)$, means that {\em all 
possible distinct equivalence classes $(I)$} are summed.  Note that each
equivalence class appears only {\em once} in the sum.  In all 
cases of interest to us, it turns out that there are  only a finite number of $(I)$'s such 
that $f(I) \neq 0$.

Now we can introduce the formula that defines the following summation:
\begin{equation}
   \sum_{I_1 I_2 \cdots I_n = (I)} X^{I_1, I_2, \ldots, I_n} \equiv
   \sum_{I_1, I_2, \ldots, I_n} \delta^{I_1 I_2 \cdots I_n}_{(I)}
   X^{I_1, I_2, \ldots, I_n}.
\label{1-1.3}
\end{equation}   
In words, in Eq.(\ref{1-1.3}) we sum over all distinct ways of cyclically 
permuting $I$, and then all {\em distinct sets} of $n$ non-empty sequences 
$I_1$, $I_2$, \ldots, $I_n$ (but note that within a particular
set of $I_1$, $I_2$, \ldots, $I_n$, some of the sequences can be identical) such that $I_1 
I_2 \cdots I_n$ is the same as this permuted sequence.  This equation, together with Eq.(\ref{1-1.2.1}), then leads 
to
\begin{equation}
   \delta^{I_1 I_2 \cdots I_n}_{(I)} = \sum_{I'_1 I'_2 \cdots I'_n = (I)}
   \delta^{I_1}_{I'_1} \delta^{I_2}_{I'_2} \cdots \delta^{I_n}_{I'_n}.
\label{1-1.5}
\end{equation}
A direct consequence of Eq.(\ref{1-1.5}) is
\[ \delta^{I_1 I_2 \cdots I_n}_{(I)} = \delta^{I_2 I_3 \cdots I_n I_1}_{(I)}. \]   
In addition, the reader can verify from 
Eq.(\ref{1-1.3}) together with Eqs.(\ref{1-1.1}), (\ref{1-1.2}) and (\ref{1-1.2.1}) that
\begin{eqnarray}
   \sum_{I_1 I_2 \cdots I_n = (I)} & = & \sum_{I_1, I_2, \ldots, I_n}
   \left( \delta^{I_1 I_2 \cdots I_n}_I + \sum_{I_{11} I_{12} = I_1} 
   \delta^{I_{12} I_2 I_3 \cdots I_n I_{11}}_I + 
   \delta^{I_2 I_3 \cdots I_n I_1}_I \right. \nonumber \\
   & & + \sum_{I_{21} I_{22} = I_2}
   \delta^{I_{22} I_3 \cdots I_n I_1 I_{21}}_I + \cdots + 
   \delta^{I_n I_1 I_2 \cdots I_{n-1}}_I \nonumber \\
   & & \left. + \sum_{I_{n1} I_{n2} = I_n}
   \delta^{I_{n2} I_1 I_2 \cdots I_{n-1} I_{n1}}_I \right) .
\label{1-1.4}
\end{eqnarray}
   
\section{Linear Independence of String-Like Operators}
\label{a1}

In this appendix, we are going to show that the set of all $\f^{(I)}_{(J)}$'s defined by Eq.(\ref{5.3}) and
all $\gamma^I_J$'s defined by Eqs.(\ref{5.1}) and (\ref{8.1}) is linearly independent.  This can be proved by
{\em ad absurdum} as follows.  Assume on the contrary that the set is linearly dependent.  Then there exists
an equation
\begin{equation}
    \sum_{p=1}^r \alpha_p \f^{(I_p)}_{(J_p)} + \sum_{p=r+1}^s \alpha_p \gamma^{I_p}_{J_p} = 0
\label{a1.1}
\end{equation}
such that $r$ and $s$ are positive integers with $r \leq s$ ($r=s$ means that the second sum vanishes), and
all $\alpha_p$'s for $p = 1$, 2, \ldots, and $s$ are non-zero complex constants.  In addition, in the above 
equation either $(I_p) \neq (I_q)$ or $(J_p) \neq (J_q)$ if $1 \leq p \leq r$, $1 \leq q \leq r$ and $p \neq q$, 
and either $I_p \neq I_q$ or $J_p \neq J_q$ if $r+1 \leq p \leq s$, $r+1 \leq q \leq s$ and $p \neq q$.  

Assume that $s > r$.  We can assume without loss of generality that 
\begin{enumerate}
\item $J_{r+1} = J_{r+2} = \cdots = J_{r+x}$;
\item $J_{r+1} \neq J_{r+x+1}$, $J_{r+1} \neq J_{r+x+2}$, \ldots, and $J_{r+1} \neq J_s$; and
\item $\#(J_{r+1}) \leq \#(J_p)$ for all $p = r+1, r+2$, \ldots, and $s$
\end{enumerate}
for some integer $x$ such that $r < x \leq s$.
Consider the action of the L.H.S. of Eq.(\ref{a1.1}) on $s^{J_{r+1}}$.  We get
\begin{equation}
   \left( \sum_{p=1}^r \alpha_p \f^{(I_p)}_{(J_p)} + \sum_{p=r+1}^s \alpha_p \gamma^{I_p}_{J_p} \right)
   s^{J_{r+1}} = \sum_{p=r+1}^{r+x} \alpha_p s^{I_p}.
\label{a1.2}
\end{equation}
Combining Eqs.(\ref{a1.1}) and (\ref{a1.2}) yields
$$ \sum_{p=r+1}^{r+x} \alpha_p s^{I_p} = 0, $$
which is impossible because $s^{I_r}$, $s^{I_{r+1}}$, \ldots, $s^{I_{r+x}}$ are linearly
independent.  Therefore, $s = r$ and Eq.(\ref{a1.1}) can be simplified to
\begin{equation}
   \sum_{p=1}^r \alpha_p \f^{(I_p)}_{(J_p)} = 0.
\label{a1.3}
\end{equation}
Again we can assume without loss of generality that
\begin{enumerate}
\item $(J_1) = (J_2) = \cdots = (J_y)$; and
\item $(J_1) \neq (J_{y+1})$, $(J_1) \neq (J_{y+2}), \ldots,$ and $(J_1) \neq (J_r)$
\end{enumerate}
for some integer $y$ such that $1 \leq y \leq r$.  Consider the action of the L.H.S. of Eq.(\ref{a1.3}) on 
$\f^{(I_1)}_{(J_1)}$:
\[ \left( \sum_{p=1}^r \alpha_p \f^{(I_p)}_{(J_p)} \right) \P^{(I_1)}_{(J_1)} = 
   \sum_{p=1}^y \delta^{J_1}_{(J_1)} \alpha_p \P^{(I_p)}. \]
However, the R.H.S. of this equation is impossible to vanish because $\P^{(I_1)}$, $\P^{(I_2)}$, \ldots,
$\P^{(I_y)}$ are linearly independent.  Thus the set of all $\f^{(I)}_{(J)}$'s together with all $\gamma^I_J$'s
is linearly independent.  Q.E.D.

\section{Multiplication of Two String-Like Operators}
\label{a2}

The assertion that the product $\gamma^I_J \gamma^K_L$ cannot be written in general as a finite linear combination
of $\gamma$'s and $\tilde{f}$'s can be proved by contradiction as follows.  Consider
the case when $\Lambda = 1$.  Let $\gamma^a_b = \gamma^{11\ldots 1}_{11\ldots 1}$, where the number 1 shows up
$a$ times in the superscript and $b$ times in the subscript of $\gamma$.  Moreover, let $\P^{(c)} =
\P^{(11\ldots 1)}$, where the number 1 shows up $c$ times, and $s^d = s^{11\ldots 1}$, where the number 1 shows up
$d$ times.  

Assume that $\gamma^1_1 \gamma^1_1 = \sum_{p=1}^r \alpha_p \gamma^p_p + \sum_{q=1}^s \beta_q 
\f^{(q)}_{(q)}$, where $\alpha_1$, $\alpha_2$, \ldots, $\alpha_r$, $\beta_1$, $\beta_2$, \ldots, and $\beta_s$ are 
non-zero complex numbers for some positive integers $r$ and $s$.  Then from the equations $\gamma^1_1 \gamma^1_1
(s^1) = \gamma^1_1 (\gamma^1_1 s^1) = 1^2 s^1$, $\gamma^1_1 \gamma^1_1 (s^2) = \gamma^1_1 (\gamma^1_1 s^1) = 
2^2 s^1$, \ldots, and $\gamma^1_1 \gamma^1_1 (s^r) = \gamma^1_1 (\gamma^1_1 s^1) = r^2 s^1$, we deduce that
$\alpha_1 = 1$ and $\alpha_2 = \alpha_3 = \cdots = \alpha_r = 2$.  Hence $\gamma^1_1 \gamma^1_1 = \gamma^1_1 + 
2 \sum_{p=2}^r \gamma^p_p + \sum_{q=1}^s \beta_q \tilde{f}^{(q)}_{(q)}$.  However, $\gamma^1_1 \gamma^1_1 (s^{r+1})
= (r + 1)^2 s^{r+1}$ and $\gamma^1_1 + 2 \sum_{p=2}^r \gamma^p_p + \sum_{q=1}^s \beta_q \tilde{f}^{(q)}_{(q)} (s^{r+1})
= (r^2 + 2 r - 1) s^{r+1}$, leading to a contradiction.

Thus we assume instead $\gamma^1_1 \gamma^1_1 = \sum_{q=1}^s \beta_q \f^{(q)}_{(q)}$ where the $\beta_q$'s are
non-zero complex numbers.  However, $\gamma^1_1 \gamma^1_1 (\P^{(s+1)}) = (s+1)^2 \P^{(s+1)}$ whereas 
$\sum_{q=1}^s \beta_q \f^{(q)}_{(q)} \P^{(s+1)} = 0$, leading to a contradiction, too.  Consequently, it is 
impossible to write $\gamma^1_1 \gamma^1_1$ as a finite linear combination of $\gamma$'s and $\tilde{f}$'s.

This proof can be easily generalized to the case $\Lambda > 1$.  Q.E.D. 

\section{The Commutation Relations of the String-Like Operators}
\label{a4}
    			   	   
What we need to do is to show that the three equations satisfy
\begin{eqnarray}
   \lbrack \g^I_J, \g^K_L \rbrack \P^{(P)} & = & \g^I_J (\g^K_L \P^{(P)}) - \g^K_L (\g^I_J \P^{(P)});
\label{a4.0.1} \\
   \lbrack \g^I_J, \f^{(K)}_{(L)} \rbrack \P^{(P)} & = & \g^I_J (\f^{(K)}_{(L)} \P^{(P)}) - 
   \f^{(K)}_{(L)} (\g^I_J \P^{(P)});
\label{a4.0.2} \\
   \lbrack \f^{(I)}_{(J)}, \f^{(K)}_{(L)} \rbrack \P^{(P)} & = & \f^{(I)}_{(J)} (\f^{(K)}_{(L)} \P^{(P)}) 
   - \f^{(K)}_{(L)} (\f^{(I)}_{(J)} \P^{(P)});
\label{a4.0.3} \\
   \lbrack \g^I_J, \g^K_L \rbrack s^M & = & \g^I_J (\g^K_L s^M) - \g^K_L (\g^I_J s^M);
\label{a4.0.4} \\
   \lbrack \g^I_J, \f^{(K)}_{(L)} \rbrack s^M & = & \g^I_J (\f^{(K)}_{(L)} s^M) - \f^{(K)}_{(L)} (\g^I_J s^M); \; 
   \mbox{and}
\label{a4.0.5} \\
   \lbrack \f^{(I)}_{(J)}, \f^{(K)}_{(L)} \rbrack s^M & = & \f^{(I)}_{(J)} (\f^{(K)}_{(L)} s^M) - 
   \f^{(K)}_{(L)} (\f^{(I)}_{(J)} s^M).
\label{a4.0.6}
\eeq
for any integer sequences $I$, $J$, $K$, $L$, $M$ and $P$.  Eqs.(\ref{a4.0.5}) and (\ref{a4.0.6}) are trivially
true.  That Eq.(\ref{5.7}) satisfies Eq.(\ref{a4.0.3}) is also straightforward.  What is remaining is whether 
Eq.(\ref{5.6}) satisfies Eq.(\ref{a4.0.2}), and Eq.(\ref{5.5}) satisfies Eqs.(\ref{a4.0.4}) and (\ref{a4.0.1}).

Consider Eq.(\ref{5.6}).  The action of the Lie bracket operator on the L.H.S. of this 
equation on $\P^{(P)}$, where $P$ is arbitrary, can be evaluated using  
Eqs.(\ref{5.1.1}) and (\ref{5.3}), and we get
\begin{eqnarray}
   \lbrack \g^I_J, \f^{(K)}_{(L)} \rbrack \P^{(P)} & = & \sum_{(Q)}      
   \left( \delta^I_{(Q)} \delta^K_{(J)} \delta^P_{(L)} +
   \sum_A \delta^{I A}_{(Q)} \delta^K_{(J A)} \delta^P_{(L)} \right. \nonumber \\ 
   & & - \left. \delta^K_{(Q)} \delta^I_{(L)} \delta^P_{(J)}
   - \sum_{A'} \delta^K_{(Q)} \delta^{I A'}_{(L)} \delta^P_{(J A')} \right)       
   \P^{(Q)}.
\label{a4.1}
\end{eqnarray}
On the other hand, the action of the operators on the R.H.S. of Eq.(\ref{5.6}) 
(let us call this linear combination of operators $g\f ^{I, K}_{J, L}$) on 
$\P^{(P)}$ is 
\begin{eqnarray}
   g\f ^{I, K}_{J, L} \P^{(P)} & = & \sum_{(Q)} \left( \delta^K_{(J)} \delta^I_{(Q)} \delta^P_{(L)} 
   + \sum_{K_1 K_2 = (K)} \delta^{K_1}_J \delta^{I K_2}_{(Q)} \delta^P_{(L)}
   \right. \nonumber \\
   & & \left. - \delta^I_{(L)} \delta^K_{(Q)} \delta^P_{(J)} - \sum_{L_1 L_2 = (L)}   
   \delta^I_{L_2} \delta^K_{(Q)} \delta^P_{(L_1 J)} \right) \P^{(Q)}.  
\label{a4.2}
\end{eqnarray}
The R.H.S of Eqs.(\ref{a4.1}) and (\ref{a4.2}) can be seen to the same by using 
the delta function defined in Eq.(\ref{1-1.2}).

Let us determine the correctness of Eq.(\ref{5.5}).  The properties of the delta functions discussed in 
Appendix~\ref{s1-1} will be extensively used.  To verify Eq.(\ref{a4.0.4}), consider the action of the Lie bracket 
operator on the L.H.S. of Eq.(\ref{5.5}) on $s^P$, where $P$ is arbitrary:  
\begin{eqnarray}
   \lefteqn{ \left[ \gamma^I_J, \gamma^K_L \right] s^P = \sum_Q \left(
   \delta^I_Q \delta^K_J \delta^P_L + \sum_C \delta^I_Q \delta^{C K}_J
     \delta^P_{C L} + \sum_D \delta^I_Q \delta^{K D}_J \delta^P_{L D} \right. }
     \nonumber \\
   & & + \sum_{C, D} \delta^I_Q \delta^{C K D}_J \delta^P_{C L D} +
   \sum_A \delta^{A I}_Q \delta^K_{A J} \delta^P_L +
   \sum_{A, C} \delta^{A I}_Q \delta^{C K}_{A J} \delta^P_{C L} \nonumber \\
   & & + \sum_{A, D} \delta^{A I}_Q \delta^{K D}_{A J} \delta^P_{L D} +
   \sum_{A, C, D} \delta^{A I}_Q \delta^{C K D}_{A J} \delta^P_{C L D}
   + \sum_B \delta^{I B}_Q \delta^K_{J B} \delta^P_L  \nonumber \\
   & & + \sum_{B, C} \delta^{I B}_Q \delta^{C K}_{J B} \delta^P_{C L} + 
   \sum_{B, D} \delta^{I B}_Q \delta^{K D}_{J B} \delta^P_{L D} +
   \sum_{B, C, D} \delta^{I B}_Q \delta^{C K D}_{J B} \delta^P_{C L D}
   \nonumber \\
   & & + \sum_{A, B} \delta^{A I B}_Q \delta^K_{A J B} \delta^P_L +
   \sum_{A, B, C} \delta^{A I B}_Q \delta^{C K}_{A J B} \delta^P_{C L} +
   \sum_{A, B, D} \delta^{A I B}_Q \delta^{K D}_{A J B} \delta^P_{L D}
   \nonumber \\
   & & + \left. \sum_{A, B, C, D} \delta^{A I B}_Q \delta^{C K D}_{A J B} 
   \delta^P_{C L D} \right) s^Q - (I \leftrightarrow K, J \leftrightarrow L).
\label{a4.9}
\end{eqnarray}
After a tedious calculation (for more details on the intermediate steps, please see Ref.~\cite{lee}), 
Eq.(\ref{a4.9}) leads to
\begin{eqnarray}
   \lefteqn{ \left[ \gamma^I_J, \gamma^K_L \right] s^P = } \nonumber \\   
   & & \sum_Q \left\{ \delta^K_J ( \delta^I_Q \delta^P_L 
   + \sum_E \delta^{E I}_Q \delta^P_{E L}
   + \sum_F \delta^{I F}_Q \delta^P_{L F} + 
   \sum_{E, F} \delta^{E I F}_Q \delta^P_{E L F} ) \right. \nonumber \\
   & & + \sum_{J_1 J_2 = J} \delta^K_{J_2} (\delta^I_Q \delta^P_{J_1 L} 
   + \sum_E \delta^{E I}_Q \delta^P_{E J_1 L} + 
   \sum_F \delta^{I F}_Q \delta^P_{J_1 L F} \nonumber \\
   & & + \sum_{E, F} \delta^{E I F}_Q \delta^P_{E J_1 L F} )
   + \sum_{K_1 K_2 = K} \delta^{K_1}_J (\delta^{I K_2}_Q \delta^P_L
   + \sum_E \delta^{E I K_2}_Q \delta^P_{E L} \nonumber \\
   & & + \sum_F \delta^{I K_2 F}_Q \delta^P_{L F} +
   \sum_{E, F} \delta^{E I K_2 F}_Q \delta^P_{E L F} )
   + \sum_{J_1 J_2 = J} \delta^K_{J_1} (\delta^I_Q \delta^P_{L J_2} \nonumber \\
   & & + \sum_E \delta^{E I}_Q \delta^P_{E L J_2} +
   \sum_F \delta^{I F}_Q \delta^P_{L J_2 F} +
   \sum_{E, F} \delta^{E I F}_Q \delta^P_{E L J_2 F} ) \nonumber \\
   & & + \sum_{K_1 K_2 = K} \delta^{K_2}_J (\delta^{K_1 I}_Q \delta^P_L
   + \sum_E \delta^{E K_1 I}_Q \delta^P_{E L} +
   \sum_F \delta^{K_1 I F}_Q \delta^P_{L F} \nonumber \\
   & & + \sum_{E, F} \delta^{E K_1 I F}_Q \delta^P_{E L F} )
   + \sum_{\begin{array}{l}
   	      J_1 J_2 = J \\
   	      K_1 K_2 = K
   	   \end{array}} \delta^{K_1}_{J_2}
   (\delta^{I K_2}_Q \delta^P_{J_1 L} + 
   \sum_E \delta^{E I K_2}_Q \delta^P_{E J_1 L} \nonumber \\
   & & + \sum_F \delta^{I K_2 F}_Q \delta^P_{J_1 L F} +
   \sum_{E, F} \delta^{E I K_2 F}_Q \delta^P_{E J_1 L F} )
   + \sum_{\begin{array}{l}
   	      J_1 J_2 = J \\
   	      K_1 K_2 = K
   	   \end{array}} \delta^{K_2}_{J_1}
   (\delta^{K_1 I}_Q \delta^P_{L J_2} \nonumber \\
   & & + \sum_E \delta^{E K_1 I}_Q \delta^P_{E L J_2} +
   \sum_F \delta^{K_1 I F}_Q \delta^P_{L J_2 F} +
   \sum_{E, F} \delta^{E K_1 I F}_Q \delta^P_{E L J_2 F} ) \nonumber \\
   & & + \sum_{J_1 J_2 J_3 = J} \delta^K_{J_2}
   (\delta^I_Q \delta^P_{J_1 L J_3} +
   \sum_E \delta^{E I}_Q \delta^P_{E J_1 L J_3} \nonumber \\
   & & + \sum_F \delta^{I F}_Q \delta^P_{J_1 L J_3 F} +
   \sum_{E, F} \delta^{E I F} \delta^P_{E J_1 L J_3 F} )
   + \sum_{K_1 K_2 K_3 = K} \delta^{K_2}_J
   (\delta^{K_1 I K_3}_Q \delta^P_L \nonumber \\ 
   & & + \sum_E \delta^{E K_1 I K_3}_Q \delta^P_{E L} +
   \sum_F \delta^{K_1 I K_3 F}_Q \delta^P_{L F} +
   \sum_{E, F} \delta^{E K_1 I K_3 F}_Q \delta^P_{E L F} ) \nonumber \\
   & & \left. + \Gamma_1 \right\} s^Q - (I \leftrightarrow K, J \leftrightarrow L)
\label{a4.10}
\end{eqnarray}    
where
\begin{eqnarray}
   \Gamma_1 & = & \delta^{I K}_Q \delta^P_{J L} + 
   \sum_E \delta^{E I K}_Q \delta^P_{E J L} +
   \sum_F \delta^{I F K}_Q \delta^P_{J F L} +
   \sum_G \delta^{I K G}_Q \delta^P_{J L G} \nonumber \\
   & & + \sum_{E, F} \delta^{E I F K}_Q \delta^P_{E J F L} +
   \sum_{F, G} \delta^{I F K G}_Q \delta^P_{J F L G} +
   \sum_{E, G} \delta^{E I K G}_Q \delta^P_{E J L G} \nonumber \\
   & & + \sum_{E, F, G} \delta^{E I F K G}_Q \delta^P_{E J F L G}
   + (I \leftrightarrow K, J \leftrightarrow L)  
\label{a4.11}
\end{eqnarray}
If we substitute Eq.(\ref{a4.10}) without $\Gamma_1$ into Eq.(\ref{a4.9}), we will 
obtain exactly the action of operators on the R.H.S. of Eq.(\ref{5.6}) on $s^P$.  
$\Gamma_1$ is reproduced when $I$ and $K$ are interchanged with $J$ and $L$ 
respectively in Eq.(\ref{a4.11}) and so it is cancelled.  Consequently, 
Eq.(\ref{a4.0.4}) is indeed satisfied.

Now let us verify Eq.(\ref{a4.0.1}).  Consider the Lie bracket operator on the L.H.S. of Eq.(\ref{5.5}) on  
$\P^{(P)}$, where $P$ is again arbitrary.  We then obtain
\begin{eqnarray}
   \left[ \g^I_J, \g^K_L \right] \P^{(P)} & = & \sum_{(Q)} \left(
   \delta^I_{(Q)} \delta^K_{(J)} \delta^P_{(L)} + \sum_A \delta^{I A}_{(Q)}
   \delta^K_{(J A)} \delta^P_{(L)} + \sum_B \delta^I_{(Q)} \delta^{K B}_{(J)}
   \delta^P_{(L B)} \right. \nonumber \\
   & & \left. + \sum_{A, B} \delta^{I A}_{(Q)} \delta^{K B}_{(J A)} 
   \delta^P_{(L B)} \right) \P^{(Q)} - 
   (I \leftrightarrow K, J \leftrightarrow L).
\label{a4.3}
\end{eqnarray}
The first three summations on the R.H.S. of this equation can be turned into the 
following expressions:
\begin{eqnarray}
   \delta^I_{(Q)} \delta^K_{(J)} \delta^P_{(L)} & = & 
   \delta^K_J \delta^I_{(Q)} \delta^P_{(L)} +
   \sum_{\begin{array}{l}
   	    J_1 J_2 = J \\
   	    K_1 K_2 = K
   	 \end{array}}
   \delta^{K_1}_{J_2} \delta^{K_2}_{J_1} \delta^I_{(Q)} \delta^P_{(L)};
\label{a4.4} \\
   \sum_A \delta^{I A}_{(Q)} \delta^K_{(J A)} \delta^P_{(L)} & = &
   \sum_{K_1 K_2 = K} \delta^{K_1}_J \delta^{I K_2}_{(Q)} \delta^P_{(L)} +
   \sum_{K_1 K_2 = K} \delta^{K_2}_J \delta^{I K_1}_{(Q)} \delta^P_{(L)}
   \nonumber \\
   & & + \sum_{K_1 K_2 K_3 = K} \delta^{K_2}_J \delta^{K_1 I K_3}_{(Q)}
   \delta^P_{(L)} \nonumber \\
   & & + \sum_{\begin{array}{l}
   		  J_1 J_2 = J \\
   		  K_1 K_2 K_3 = K
   	       \end{array}}
   \delta^{K_1}_{J_2} \delta^{K_3}_{J_1} \delta^{I K_2}_{(Q)} \delta^L_{(L)}
   \mbox{; and}	    
\label{a4.5} \\
   \sum_B \delta^I_{(Q)} \delta^{K B}_{(J)} \delta^P_{(L B)} & = &
   \sum_{J_1 J_2 = J} \delta^K_{J_2} \delta^I_{(Q)} \delta^P_{(J_1 L)} +
   \sum_{J_1 J_2 = J} \delta^K_{J_1} \delta^I_{(Q)} \delta^P_{(J_2 L)} 
   \nonumber \\
   & & + \sum_{J_1 J_2 J_3 = J} \delta^K_{J_2} \delta^I_{(Q)}
   \delta^P_{(J_1 L J_3)} \nonumber \\
   & & + \sum_{\begin{array}{l}
   		  J_1 J_2 J_3 = J \\
   		  K_1 K_2 = K
   	       \end{array}}
   \delta^{K_1}_{J_3} \delta^{K_2}_{J_1} \delta^I_{(Q)} \delta^P_{(J_2 L)}.
\label{a4.6}
\end{eqnarray}
The fourth summation is more complicated.  This can be manipulated to be:
\begin{eqnarray}
   \lefteqn{\sum_{A, B} \delta^{I A}_{(Q)} \delta^{K B}_{(J A)} \delta^P_{(L B)} 
   = \delta^K_J \sum_C \delta^{I C}_{(Q)} \delta^P_{(L C)} +
   \sum_{J_1 J_2 = J} \sum_C \delta^K_{J_1} \delta^{I C}_{(Q)}
   \delta^P_{(L J_2 C)} } \nonumber \\
   & & + \sum_{K_1 K_2 = K} \sum_C \delta^{K_1}_J \delta^{I K_2 C}_{(Q)} 
   \delta^P_{(L C)} +
   \sum_{J_1 J_2 = J} \sum_C \delta^K_{J_2} \delta^{I C}_{(Q)} 
   \delta^P_{(J_1 L C)} \nonumber \\
   & & + \sum_{K_1 K_2 = K} \sum_C \delta^{K_2}_J \delta^{K_1 I C}_{(Q)}
   \delta^P_{(L C)} +
   \sum_{\begin{array}{l}
   		J_1 J_2 = J \\
   		K_1 K_2 = K
   	     \end{array}}
   \delta^{K_2}_{J_1} \delta^{I K_1}_{(Q)} \delta^P_{(J_2 L)} \nonumber \\	     	      
   & & + \sum_{\begin{array}{l}
   		  J_1 J_2 = J \\
  		  K_1 K_2 = K
  	       \end{array}}
   \delta^{K_1}_{J_2} \delta^{I K_2}_{(Q)} \delta^P_{(J_1 L)} +
   \sum_{\begin{array}{l}
	    J_1 J_2 = J \\
	    K_1 K_2 = K
	 \end{array}} \sum_C
   \delta^{K_1}_{J_2} \delta^{I K_2 C}_{(Q)} \delta^P_{(J_1 L C)} \nonumber \\
   & & + \sum_{\begin{array}{l}
		  J_1 J_2 = J \\
		  K_1 K_2 = K
	       \end{array}} \sum_C
   \delta^{K_2}_{J_1} \delta^{K_1 I C}_{(Q)} \delta^P_{(L J_2 C)} \nonumber \\
   & & + \sum_{J_1 J_2 J_3 = J} \sum_C
   \delta^K_{J_2} \delta^{I C}_{(Q)} \delta^P_{(J_1 L J_3 C)} \nonumber \\
   & & + \sum_{K_1 K_2 K_3 = K} \sum_C
   \delta^{K_2}_J \delta^{K_1 I K_3 C}_{(Q)} \delta^P_{(L C)} \nonumber \\
   & & + \sum_{(Q)} \sum_{\begin{array}{l}
			     J_1 J_2 J_3 = J \\
			     K_1 K_2 K_3 = K
			  \end{array}}
   \delta^{K_1}_{J_3} \delta^{K_3}_{J_1} \delta^{I K_2}_{(Q)} \delta^P_{(J_2 L)}  
   + \Gamma_2
\label{a4.7}
\end{eqnarray}
where
\begin{eqnarray}
   \Gamma_2 & = & \sum_{\begin{array}{l}
   			 P_1 P_2 = (P) \\
   			 Q_1 Q_2 = (Q)
   		      \end{array}}
   \delta^I_{Q_1} \delta^K_{Q_2} \delta^{P_1}_L \delta^{P_2}_J +
   \sum_{\begin{array}{l}
   	    P_1 P_2 P_3 = (P) \\
   	    Q_1 Q_2 Q_3 = (Q)
   	 \end{array}}
   \delta^K_{Q_1} \delta^I_{Q_2} \delta^{P_1}_{Q_3} \delta^{P_2}_L 
   \delta^{P_3}_J \nonumber \\
   & & + \sum_{\begin{array}{l}
   		  P_1 P_2 P_3 = (P) \\
   		  Q_1 Q_2 Q_3 = (Q)
  	       \end{array}}		      	 
   \delta^I_{Q_1} \delta^K_{Q_2} \delta^{P_1}_{Q_3} \delta^{P_2}_J 
   \delta^{P_3}_L \nonumber \\
   & & + \sum_{\begin{array}{l}
   		  P_1 P_2 P_3 P_4 = (P) \\
   		  Q_1 Q_2 Q_3 Q_4 = (Q)
  	       \end{array}}
   \delta^I_{Q_1} \delta^K_{Q_3} \delta^{P_1}_J \delta^{P_3}_L 
   \delta^{P_2}_{Q_2} \delta^{P_4}_{Q_4}.
\label{a4.8}
\end{eqnarray}

If we substitute Eqs.(\ref{a4.4}), (\ref{a4.5}), (\ref{a4.6}) and (\ref{a4.7}) 
without $\Gamma_2$ into Eq.(\ref{a4.3}), we will  obtain exactly the action of the 
operators on the R.H.S. of Eq.(\ref{5.5}) on $\P^{(P)}$.  $\Gamma_2$ is reproduced 
when $I$ and $K$ are interchanged with $J$ and $L$ respectively in 
Eq.(\ref{a4.8}) and so it is cancelled.  Hence Eq.(\ref{a4.0.1}) is also satisfied.  Consequently, Eq.(\ref{5.5}) 
is true. 
Q.E.D. 

\section{Cartan Subalgebra of the Heterix Algebra}
\label{a8}

First of all, it is obvious that ${\cal H}$ is abelian.  {\em A fortiori}, ${\cal H}$ is nilpotent.  To proceed on,
let us digress and state the following two lemmas:
\begin{lemma}
Let
\begin{eqnarray*}
   \left[ \g^I_I, \g^K_L \right] & = & \sum_{k=1}^p \alpha^{N_k}_{M_k} \g^{M_k}_{N_k} +
				       \sum_{k=p+1}^q \alpha^{N_k}_{M_k} \f^{(M_k)}_{(N_k)}; \\
   \left[ \g^I_I, \f^{(K)}_{(L)} \right] & = & \sum_{k=1}^q \alpha^{N_k}_{M_k} \f^{(M_k)}_{(N_k)}; \\
   \left[ \f^{(I)}_{(I)}, \g^K_L \right] & = & \sum_{k=1}^q \alpha^{N_k}_{M_k} \f^{(M_k)}_{(N_k)} \mbox{; and} \\
   \left[ \f^{(I)}_{(I)}, \f^{(K)}_{(L)} \right] & = & ( \delta^K_{(I)} - \delta^I_{(L)} ) \f^{(K)}_{(L)}
\end{eqnarray*}
where $p$ and $q$ are finite non-negative integers such that $q \geq p$, $M_k$'s and $N_k$'s are positive integer 
sequences such that $\g^{M_k}_{N_k} \neq \g^{M_k'}_{N_k'}$ and $\f^{(M_k)}_{(N_k)} \neq \f^{(M_k')}_{(N_k')}$ for 
$k \neq k'$, and $\alpha^{N_k}_{M_k}$'s are non-zero numerical coefficients.  Then
\[ \#(M_k) - \#(N_k) = \#(K) - \#(L) \]
for every $k = 1, 2, \ldots, p$ or $q$.
\label{la8.1}
\end{lemma}
This lemma can be proved by using Eqs.(\ref{5.5}), (\ref{5.6}) and (\ref{5.7}) with $J = I$.
\begin{lemma}
With the same assumptions as in the previous lemma, we have
\[ \#(M_k) + \#(N_k) \geq \#(K) + \#(L) \]
for every $k = 1, 2, \ldots, p$ or $q$.
\label{la8.2}
\end{lemma}
This lemma can also be proved by using the same equations with $J = I$.  

Let $r$ and $s$ be {\em positive} integers such that $s \geq r$.  We are now ready to show for arbitrary 
non-zero complex numbers $\xi^{L_i}_{K_i}$'s where $i = 1, 2, \ldots, s$ and arbitrary integer sequences $L_i$'s 
and $K_i$'s such that $\g^{K_i}_{L_i} \neq \g^{K_i'}_{L_i'}$ and $\f^{(K_i)}_{(L_i)} \neq \f^{(K_i')}_{(L_i')}$ for 
$i \neq i'$, and $K_i \neq L_i$ for at least one $i$ in $\{ 1, 2, \ldots, r \}$ that there exisits a sequence $I$ 
such that
\[ \lbrack \g^I_I, \sum_{i=1}^r \xi^{L_i}_{K_i} \g^{K_i}_{L_i} + \sum_{i=r+1}^s \xi^{L_i}_{K_i} \f^{(K_i)}_{(L_i)} 
   \rbrack \]
does not belong to ${\cal H}$.  Indeed, let $j$ be an integer such that
\begin{enumerate}
\item $K_j \neq L_j$;
\item $\#(K_j) - \#(L_j) \geq \#(K_i) - \#(L_i)$ for all $i = 1, 2, \ldots$ and $r$; and
\item $\#(K_j) + \#(L_j) \leq \#(K_i) + \#(L_i)$ for any $i = 1, 2, \ldots$ or $r$ such that 
      $\#(K_j) - \#(L_j) = \#(K_i) - \#(L_i)$.
\end{enumerate}
If $\#(K_j) \geq \#(L_j)$, then consider
\begin{eqnarray}
   \lefteqn{ \left[ \g^{K_j}_{K_j}, \sum_{i=1}^r \xi^{L_i}_{K_i} \g^{K_i}_{L_i} + 
   \sum_{i=r+1}^s \xi^{L_i}_{K_i} \f^{(K_i)}_{(L_i)} \right] = } \nonumber \\
   & &  \xi^{L_j}_{K_j} \g^{K_j}_{L_j} + 
   \xi^{L_j}_{K_j} \left( \sum_{k=2}^{p_j} \alpha^{N_{jk}}_{M_{jk}} \g^{M_{jk}}_{N_{jk}} +
   \sum_{k = p_j + 1}^{q_j} \alpha^{N_{jk}}_{M_{jk}} \f^{(M_{jk})}_{(N_{jk})} \right) \nonumber \\
   & & + \sum_{\begin{array}{l} i=1 \\ i \neq j \end{array}}^r 
   \xi^{L_i}_{K_i} \left(\sum_{k=1}^{p_i} \alpha^{N_{ik}}_{M_{ik}} \g^{M_{ik}}_{N_{ik}} +
   \sum_{k = p_i + 1}^{q_i} \alpha^{N_{ik}}_{M_{ik}} \f^{(M_{ik})}_{(N_{ik})} \right) \nonumber \\
   & & + \sum_{i=r+1}^s \xi^{L_i}_{K_i} \sum_{k=1}^{q'_i} \alpha^{N_{ik}}_{M_{ik}} \f^{(M_{ik})}_{(N_{ik})}
\label{a8.1}
\end{eqnarray}
where we have set $\alpha^{N_{j1}}_{M_{j1}} = 1$, and each $p_i$, $q_i$ for $i = 1, 2, \ldots, r$ and $q'_i$ for 
$i = r + 1, 2, \ldots, s$ are dependent on $i$.  Let us assume that
\begin{equation}
   \g^{M_{ik}}_{N_{ik}} = \g^{K_i}_{L_i}
\label{a8.2}
\end{equation}
for some $i \in \{ 1, 2, \ldots, r \}$ but $i \neq j$ and $k \in \{ 1, 2, \ldots, p_i \}$.  Then $\#(M_{ik}) =
\#(K_j)$ and $\#(N_{ik}) = \#(L_j)$.  By Lemmas~\ref{la8.1} and \ref{la8.2}, we get $\#(K_i) = \#(K_j)$ and 
$\#(L_i) = \#(L_j)$.  However, we also know that $K_i \neq K_j$ or $L_i \neq L_j$ and  so there is no $k \in \{
1, 2, \ldots, p_i \}$ such that Eq.(\ref{a8.2}) holds.  This is a contradiction and so we conclude that
\[ \g^{M_{ik}}_{N_{ik}} \neq \g^{K_i}_{L_i} \]
for {\em all} $i \in \{ 1, 2, \ldots, r \}$ but $i \neq j$ and $k \in \{ 1, 2, \ldots, p_i \}$.  From 
Eq.(\ref{a8.1}), we deduce that 
\[ \lbrack \g^{K_j}_{K_j}, \sum_{i=1}^r \xi^{L_i}_{K_i} \g^{K_i}_{L_i} + 
   \sum_{i=r+1}^s \xi^{L_i}_{K_i} \f^{(K_i)}_{(L_i)} \rbrack \]
does not belong to ${\cal H}$.  Similarly, if $\#(K_j) \leq \#(L_j)$, then
\[ \lbrack \g^{L_j}_{L_j}, \sum_{i=1}^r \xi^{L_i}_{K_i} \g^{K_i}_{L_i} + 
   \sum_{i=1}^s \xi^{L_i}_{K_i} \f^{(K_i)}_{(L_i)} \rbrack \]
does not belong to ${\cal H}$.  Consequently, 
\[ \sum_{i=1}^r \xi^{L_i}_{K_i} \g^{K_i}_{L_i} + \sum_{i=r+1}^s \xi^{L_i}_{K_i} \f^{(K_i)}_{(L_i)} \]
does not belong to the normalizer (see Humphreys \cite{humphreys} for the definition of a normalizer) of ${\cal H}$
if $r \neq 0$ and $K_i \neq L_i$ for at least one $i = 1, 2, \leq r$.

Now, consider
\[ \sum_{i=1}^r \xi^{L_i}_{K_i} \g^{K_i}_{K_i} + \sum_{i=r+1}^s \xi^{L_i}_{K_i} \f^{(K_i)}_{(L_i)} \]
where $r$ is a non-negative integer and $s$ a positive integer, $\xi^{L_i}_{K_i}$'s are complex constants and
$\f^{(K_i)}_{(L_i)} \neq \f^{(K_i')}_{(L_i')}$ for $i \neq i'$, and $(K_i) \neq (L_i)$ for at least one $i \in \{
r + 1, r + 2, \ldots, s \}$.  We may well assume without loss of generality that $(K_{r+1}) \neq (L_{r+1})$.  Then
\[ \left[ \f^{(K_{r+1})}_{(K_{r+1})}, \sum_{i=1}^r \xi^{L_i}_{K_i} \g^{K_i}_{K_i} + 
   \sum_{i=r+1}^s \xi^{L_i}_{K_i} \f^{(K_i)}_{(L_i)} \right] \]
does not belong to ${\cal H}$, too.  As a result, the normalizer of ${\cal H}$ is ${\cal H}$ itself.  It is 
therefore that ${\cal H}$ is a Cartan subalgebra of the heterix algebra.  Q.E.D.

\section{Root Vectors of the Heterix Algebra}
\label{a5}

Since the set of all $\f^{(K)}_{(L)}$'s is linearly independent and the root for which $\f^{(K)}_{(L)}$ is a root
vector is distinct from the root for which $\f^{(K')}_{(L')}$, where $(K) \neq (K')$ or $(L) \neq (L')$, is a root
vector, every root vector which is a finite linear combination of $\f^{(K)}_{(L)}$'s only must be a constant 
multiple of one $\f^{(K)}_{(L)}$ only.  Hence we only need to show that if a root vector is of the form
\[ f = \sum_{P, Q} a^Q_P \g^P_Q + \sum_{(P), (Q)} b^{(Q)}_{(P)} \f^{(P)}_{(Q)} \]
where a finite but {\em non-zero} number of $a^Q_P$'s are non-zero, and a finite number of $b^{(Q)}_{(P)}$'s are
non-zero as well, then $f$ must be a constant multiple of a $f^K_L$.

Indeed, let
\begin{eqnarray*}
   \left[ \g^M_M, f \right] & = & \lambda_M f \mbox{; and} \\ 
   \left[ \f^{(M)}_{(M)}, f \right] & = & \lambda_{(M)} f
\end{eqnarray*}
where each $\lambda_M$ and each $\lambda_{(M)}$ are complex constants for any integer sequence $M$.  Since
\[ \left[ \f^{(M)}_{(M)}, f \right] \in \tsalt \]
but $f \not\in \tsalt$, $\lambda_{(M)} = 0$ for any $(M)$.  Now obtain an expression for $\lambda_M$ starting with
the equation 
\begin{equation}
   \lbrack \g^M_M, f \rbrack s^K = \lambda_M f s^K
\label{a5.1}
\end{equation}
as follows.  From Eq.(\ref{8.1}), we deduce that
\begin{eqnarray*}
   \s^P_Q s^J & = & \sum_I \left( \delta^J_Q \delta^P_I + 
   \sum_{J_1 J_2 = J} \delta^{J_2}_Q \delta^{J_1 P}_I +
   \sum_{J_1 J_2 = J} \delta^{J_1}_Q \delta^{P J_2}_I \right. \\
   & & \left. \sum_{J_1 J_2 J_3 = J} \delta^{J_2}_Q \delta^{J_1 P J_3}_I
   \right) s^I.
\end{eqnarray*}
Therefore,
\begin{eqnarray}
   \lefteqn{ \left[ \s^M_M, f \right] s^K = } \nonumber \\ 
   & & \sum_{I, P, Q} \left( \delta^M_I \delta^K_Q \delta^P_M +
   \sum_{K_1 K_2 = K} \delta^M_I \delta^{K_2}_Q \delta^{K_1 P}_M + 
   \sum_{K_1 K_2 = K} \delta^M_I \delta^{K_1}_Q \delta^{P K_2}_M \right.
     \nonumber \\
   & & + \sum_{K_1 K_2 K_3 = K} \delta^M_I \delta^{K_2}_Q \delta^{K_1 P K_3}_M
   + \sum_{I_1 I_2 = I} \delta^M_{I_2} \delta^K_Q \delta^P_I +
   \sum_{\begin{array}{l}
   	    I_1 I_2 = I \\
   	    K_1 K_2 = K
   	 \end{array}}
   \delta^M_{I_2} \delta^{K_2}_Q \delta^{K_1 P}_I \nonumber \\
   & & + \sum_{\begin{array}{l}
   	    I_1 I_2 = I \\
   	    K_1 K_2 = K
   	 \end{array}}
   \delta^M_{I_2} \delta^{K_1}_Q \delta^{P K_2}_I +
   \sum_{\begin{array}{l}
   	    I_1 I_2 = I \\
   	    K_1 K_2 K_3 = K
   	 \end{array}}
   \delta^M_{I_2} \delta^{K_2}_Q \delta^{K_1 P K_3}_I \nonumber \\
   & & + \sum_{I_1 I_2 = I} \delta^M_{I_1} \delta^K_Q \delta^P_I +
   \sum_{\begin{array}{l}
   	    I_1 I_2 = I \\
   	    K_1 K_2 = K
   	 \end{array}}
   \delta^M_{I_1} \delta^{K_2}_Q \delta^{K_1 P}_I +
   \sum_{\begin{array}{l}
   	    I_1 I_2 = I \\
   	    K_1 K_2 = K
   	 \end{array}}
   \delta^M_{I_1} \delta^{K_1}_Q \delta^{P K_2}_I \nonumber \\
   & & + \sum_{\begin{array}{l}
   		  I_1 I_2 = I \\
   		  K_1 K_2 K_3 = K
  	       \end{array}}
   \delta^M_{I_1} \delta^{K_2}_Q \delta^{K_1 P K_3}_I +
   \sum_{I_1 I_2 I_3 = I} \delta^M_{I_2} \delta^K_Q \delta^P_I \nonumber \\
   & & + \sum_{\begin{array}{l}
   		  I_1 I_2 I_3 = I \\
   		  K_1 K_2 = K
  	       \end{array}}
   \delta^M_{I_2} \delta^{K_2}_Q \delta^{K_1 P}_I
   + \sum_{\begin{array}{l}
   	      I_1 I_2 I_3 = I \\
   	      K_1 K_2 = K
  	   \end{array}}
   \delta^M_{I_2} \delta^{K_1}_Q \delta^{P K_2}_I \nonumber \\
   & & \left. + \sum_{\begin{array}{l}
   		         I_1 I_2 I_3 = I \\
   		         K_1 K_2 K_3 = K
   	              \end{array}}
   \delta^M_{I_2} \delta^{K_2}_Q \delta^{K_1 P K_3}_I \right) a^Q_P s^I
   \nonumber \\
   & & - \sum_{I, P, Q} \left( \delta^P_I \delta^K_M \delta^M_Q +
   \sum_{K_1 K_2 = K} \delta^P_I \delta^{K_2}_M \delta^K_Q + 
   \sum_{K_1 K_2 = K} \delta^P_I \delta^{K_1}_M \delta^K_Q \right.
     \nonumber \\
   & & + \sum_{K_1 K_2 K_3 = K} \delta^P_I \delta^{K_2}_M \delta^K_Q
   + \sum_{I_1 I_2 = I} \delta^P_{I_2} \delta^K_M \delta^K_{I_1 Q} +
   \sum_{\begin{array}{l}
   	    I_1 I_2 = I \\
   	    K_1 K_2 = K
   	 \end{array}}
   \delta^P_{I_2} \delta^{K_2}_M \delta^K_{I_1 Q} \nonumber \\
   & & + \sum_{\begin{array}{l}
   	    I_1 I_2 = I \\
   	    K_1 K_2 = K
   	 \end{array}}
   \delta^P_{I_2} \delta^{K_1}_M \delta^K_{I_1 Q} +
   \sum_{\begin{array}{l}
   	    I_1 I_2 = I \\
   	    K_1 K_2 K_3 = K
   	 \end{array}}
   \delta^P_{I_2} \delta^{K_2}_M \delta^K_{I_1 Q} \nonumber \\
   & & + \sum_{I_1 I_2 = I} \delta^P_{I_1} \delta^K_M \delta^K_{Q I_2} +
   \sum_{\begin{array}{l}
   	    I_1 I_2 = I \\
   	    K_1 K_2 = K
   	 \end{array}}
   \delta^P_{I_1} \delta^{K_2}_M \delta^K_{Q I_2} +
   \sum_{\begin{array}{l}
   	    I_1 I_2 = I \\
   	    K_1 K_2 = K
   	 \end{array}}
   \delta^P_{I_1} \delta^{K_1}_M \delta^K_{Q I_2} \nonumber \\
   & & + \sum_{\begin{array}{l}
   		  I_1 I_2 = I \\
   		  K_1 K_2 K_3 = K
  	       \end{array}}
   \delta^P_{I_1} \delta^{K_2}_M \delta^K_{Q I_2} +
   \sum_{I_1 I_2 I_3 = I} \delta^P_{I_2} \delta^K_M \delta^K_{I_1 Q I_3} \nonumber \\
   & & + \sum_{\begin{array}{l}
   		  I_1 I_2 I_3 = I \\
   		  K_1 K_2 = K
  	       \end{array}}
   \delta^P_{I_2} \delta^{K_2}_M \delta^K_{I_1 Q I_3}
   + \sum_{\begin{array}{l}
   	      I_1 I_2 I_3 = I \\
   	      K_1 K_2 = K
  	   \end{array}}
   \delta^P_{I_2} \delta^{K_1}_M \delta^K_{I_1 Q I_3} \nonumber \\
   & & \left. + \sum_{\begin{array}{l}
   		         I_1 I_2 I_3 = I \\
   		         K_1 K_2 K_3 = K
   	              \end{array}}
   \delta^P_{I_2} \delta^{K_2}_M \delta^K_{I_1 Q I_3} \right) a^Q_P s^I.
\label{a9.1}
\end{eqnarray}
As a result, we can combine Eqs.(\ref{a5.1}) and (\ref{a9.1}) together to obtain an equation which is too long to 
be written down here for {\em any} integer sequences $I$, $K$ and $M$.
   
Let us find an $a^S_R$ in $f$ such that $R \neq S$, $a^S_R \neq 0$,
$a^{S_1}_{R_1} = 
a^{S_2}_{R_2} = 0$ for all $R_1$'s, $S_1$'s, $R_2$'s and $S_2$'s 
such that $R_1 R_2 = R$ and $S_1 S_2 = S$, and $a^{S_2}_{R_2} = 0$ for all 
$R_2$'s and $S_2$'s such that $R_1 R_2 R_3 = R$ and $S_1 S_2 S_3 = S$ for some
$R_1$, $R_3$, $S_1$ and $S_3$.  The reader can easily convince himself or 
herself that such an $a^S_R$ always exists.  Let us choose $I = R$ and $K = S$
in 
Eq.(\ref{a9.1}).  Then when we combine Eqs.(\ref{a5.1}) and (\ref{a9.1}), we
get
\begin{eqnarray}
   \lambda_M & = & \delta^R_M + \sum_{R_1 R_2 = R} \delta^{R_1}_M 
   + \sum_{R_1 R_2 = R} \delta^{R_2}_M + \sum_{R_1 R_2 R_3 = R}
\delta^{R_2}_M
   \nonumber \\
   & & - \delta^M_S - \sum_{S_1 S_2 = S} \delta^M_{S_1}
   - \sum_{S_1 S_2 = S} \delta^M_{S_2} - \sum_{S_1 S_2 S_3 = S}
\delta^M_{S_2}.
\label{a9.3}
\end{eqnarray}
Therefore, we obtain after some manipulation that
\[ \lambda_M - \sum_{j=1}^{{\Lambda}} \lambda_{Mj} - \sum_{i=1}^{{\Lambda}}
\lambda_{iM}        
   + \sum_{i, j = 1}^{{\Lambda}} \lambda_{iMj} = \delta^R_M - \delta^M_S. \]
This means
\begin{eqnarray*}
   \left[ f^R_R , f \right] & \neq & 0 \mbox{; and} \\
   \left[ f^S_S , f \right] & \neq & 0.
\end{eqnarray*}
Thus $f \in F'_{\Lambda}$.  Since we know that every $f^K_L$ is a root vector, and the set of all $f^K_L$'s is 
linearly independent, $f = f^K_L$ for some $K$ and $L$.  Q.E.D.

\section{A Basis for a $\M = 1$ Quotient Lie Algebra}
\label{a6}

We are going to prove that the set of all $\f^{(a)}_{(b)}$, $g^1_b$ and $g^a_1$ where $a$ and $b$ are arbitrary 
positive integers form a basis for the quotient algebra of all {\em cosets} $\g^a_b$ and $\f^{(a)}_{(b)}$.  Indeed,
consider the equation
\begin{equation}
   \sum_{a,b=1}^{\infty} \alpha^{(b)}_{(a)} \f^{(a)}_{(b)} +
   \sum_{d=1}^{\infty} \alpha^d g^1_d + \sum_{c=2}^{\infty} \alpha_c g^c_1 = 0
\label{a6.1}
\end{equation}
where only a finite number of the $\alpha$'s are non-zero.  Let $n$ be an integer 
such that for all $a$'s and $b$'s such that $\alpha^{(b)}_{(a)} \neq 0$, we have 
$n > b$ and for all $d$'s such that $\alpha^d \neq 0$, we have $n > d$ also.  
Then
\begin{eqnarray*}
   \lefteqn{ \left( \sum_{a,b=1}^{\infty} \alpha^{(b)}_{(a)} \f^{(a)}_{(b)} +
   \sum_{d=1}^{\infty} \alpha^d g^1_d + \sum_{c=2}^{\infty} \alpha_c g^c_1
   \right) \P^{(n)} = } \\
   & & \sum_{d=1}^{\infty} n \alpha^d \P^{(n-d+1)} + \sum_{c=2}^{\infty}
   n \alpha_c \P^{(n+c-1)}.
\end{eqnarray*}
For the R.H.S of this equation to vanish, we need $\alpha^d = 0$ and $\alpha_c = 
0$ for all $c$'s and $d$'s.  Hence Eq.(\ref{a6.1}) becomes
\[ \sum_{a,b=1}^{\infty} \alpha^{(b)}_{(a)} \f^{(a)}_{(b)} = 0. \]
Then
\[ \sum_{a,b=1}^{\infty} \alpha^{(b)}_{(a)} \f^{(a)}_{(b)} \P^{(e)}
   = \sum_{a=1}^{\infty} e \alpha^{(e)}_{(a)} \P^{(a)} \]
for any positive integer $e$.  Thus $\alpha^{(e)}_{(a)} = 0$ also for any positive $a$ and 
$e$.  Consequently, the set of all $\f^{(a)}_{(b)}$'s, $g^1_b$'s and $g^a_1$'s 
where $a$ and $b$ are arbitrary integers is linearly independent.  Q.E.D.             

\section{Cartan Subalgebra of a $\Lambda = 1$ Quotient Lie Algebra}
\label{a7}

This can be seen as follows.  From the results of Section~\ref{s5}, we know that 
the subspace spanned by all $\f^{(a)}_{(a)}$'s and $g^1_1$ forms an abelian 
subalgebra.  
Moreover, consider a vector $v$ of the form
\[ v = \sum_{a, b = 1}^{\infty} \alpha^{(b)}_{(a)} \f^{(a)}_{(b)} +               
   \sum_{d=2}^{\infty} \alpha^d g^1_d + \sum_{c=2}^{\infty} \alpha_c g^c_1 \]
where only a finite number of the $\alpha$'s not equal to 0, and where 
$\alpha^{(a)}_{(a)} = 0$ for all $a = 1, 2, \cdots, \infty$.  If all the 
$\alpha^d$'s and $\alpha_c$'s vanish, then choose a particular $a_0$ such that there exists a $b$ with 
$\alpha^{(b)}_{(a_0)} \neq 0$.  Then 
\begin{eqnarray*}
   \left[ \f^{(a_0)}_{(a_0)}, v \right] & = & 
   \sum_{b = 1}^{\infty} a_0 \alpha^{(b)}_{(a_0)} \f^{(a_0)}_{(b)} -
   \sum_{a = 1}^{\infty} a_0 \alpha^{(a_0)}_{(a)} \f^{(a)}_{(a_0)} \\
   & \neq & 0.
\end{eqnarray*}
Hence $v$ does not commute with the subspace spanned by all $\f^{(a)}_{(a)}$'s 
and $g^1_1$.  If there exists at least one non-zero $\alpha^d$ or $\alpha_c$, set 
$m$ to be the maximum of all $a$'s and $b$'s such that $\alpha^{(b)}_{(a)} 
\neq 0$.  Then $\alpha^{(b')}_{(a')} = 0$ if either $b'$ or $a' > m$.  Use 
Eq.(\ref{6.3}) to rewrite each $g^1_d$ and $g^c_1$ such that $\alpha^d \neq 0$ 
and $\alpha_c \neq 0$ as
\[ g^1_d = \f^{(1)}_{(d)} + \f^{(2)}_{(d+1)} + \cdots + \f^{(m)}_{(m+d-1)} +
           g^{m+1}_{d+m} \]
and
\[ g^c_1 = \f^{(c)}_{(1)} + \f^{(c+1)}_{(2)} + \cdots + \f^{(m+c-1)}_{(m)} +
	   g^{c+m}_{m+1}. \]
Then
\[ v = \sum_{a, b = 1}^{\infty} \alpha'^{(b)}_{(a)} \f^{(a)}_{(b)} +
   \sum_{d=2}^{\infty} \alpha^d g^{m+1}_{d+m} + 
   \sum_{c=2}^{\infty} \alpha_c	g^{c+m}_{m+1} \]   
where 
\[ \alpha'^{(b)}_{(a)} = \left\{ \begin{array}{ll}
   \alpha^{(b)}_{(a)} + \alpha_{a-b+1} & \mbox{if $a>b$; and} \\
   \alpha^{(b)}_{(a)} + \alpha^{b-a+1} & \mbox{if $b>a$.}
   \end{array} \right. \]
It is possible for $\alpha'^{(b)}_{(a)} \neq 0$ only if $b \leq m$ or $a \leq m$. 
If $\alpha'^{(b_0)}_{(a_0)} \neq 0$ for a particular pair of numbers $b_0$ and $a_0$, then
\begin{eqnarray*}
   \left[ \f^{(b_0)}_{(b_0)}, v \right] & \neq & 0 \;\mbox{if $b_0 \leq m$; or} \\
   \left[ \f^{(a_0)}_{(a_0)}, v \right] & \neq & 0 \;\mbox{if $a_0 \leq m$}
\end{eqnarray*}
because
\begin{eqnarray*}
   \left[ \f^{(b_0)}_{(b_0)}, \sum_{d=2}^{\infty} \alpha^d g^{m+1}_{d+m} \right] & = & 0 \mbox{; and} \\
   \left[ \f^{(b_0)}_{(b_0)}, \sum_{c=2}^{\infty} \alpha_c g^{c+m}_{m+1} \right] & = & 0
\end{eqnarray*}
if $b_0 \leq m$, and
\begin{eqnarray*}
   \left[ \f^{(a_0)}_{(a_0)}, \sum_{d=2}^{\infty} \alpha^d g^{m+1}_{d+m} \right] & = & 0 \mbox{; and} \\
   \left[ \f^{(a_0)}_{(a_0)}, \sum_{c=2}^{\infty} \alpha_c g^{c+m}_{m+1} \right] & = & 0
\end{eqnarray*}
if $a_0 \leq m$.  If all $\alpha'^{(b)}_{(a)}$'s vanish, then
\[ v = \sum_{d=2}^{\infty} \alpha^d g^{m+1}_{d+m} + 
   \sum_{c=2}^{\infty} \alpha_c	g^{c+m}_{m+1} \]
and so
\[ \left[ \f^{(m+1)}_{(m+1)}, v \right] =  \sum_{d=2}^{\infty} (m + 1) \alpha^d   
   \f^{(m+1)}_{(d+m)} - \sum_{c=2}^{\infty} (m + 1) \alpha_c \f^{(c+m)}_{(m+1)} \neq 0.
\]       
Q.E.D.

\end{document}